\documentclass[twocolumn,nobibnotes, aps, pra, superscriptaddress,floatfix]{revtex4-1} 

\usepackage[normalem]{ulem} 	
\usepackage[usenames,dvipsnames]{color}
\usepackage{upgreek}
\usepackage{braket}
\usepackage{float}
\usepackage[nointegrals]{wasysym}
\usepackage{helvet}
\usepackage{hyperref}
\usepackage{xcolor}
\usepackage[english]{babel}
\usepackage{graphicx, import}
\usepackage[export]{adjustbox}
\usepackage{tikz, amsmath}
\usetikzlibrary{decorations.pathmorphing}
\tikzset{snake it/.style={decorate, decoration=snake}}
\usetikzlibrary{arrows}
\newcommand{\midarrow}{\tikz \draw[-triangle 90] (0,0) -- +(.1,0);}
\usepackage{amssymb,amsfonts,amsmath} 
\usepackage{physics}
\def \ci {\mathrm{i}}
\definecolor{pink2}{RGB}{254, 1, 154}

\usepackage[normalem]{ulem}

\begin{document}

\title{Memory effects in the density-wave imbalance in delocalized disordered systems}
\author{Paul P{\"o}pperl}
	\affiliation{\mbox{Institut für Theorie der Kondensierten Materie, Karlsruhe Institute of Technology, 76128 Karlsruhe, Germany}}
	\affiliation{\mbox{Institute for Quantum Materials and Technologies, Karlsruhe Institute of Technology, 76021 Karlsruhe, Germany}}
\author{Igor~V.~Gornyi}
	\affiliation{\mbox{Institut für Theorie der Kondensierten Materie, Karlsruhe Institute of Technology, 76128 Karlsruhe, Germany}}
	\affiliation{\mbox{Institute for Quantum Materials and Technologies, Karlsruhe Institute of Technology, 76021 Karlsruhe, Germany}}
\affiliation{Ioffe Institute, 194021 St.~Petersburg, Russia}
\author{Alexander D. Mirlin}
	\affiliation{\mbox{Institut für Theorie der Kondensierten Materie, Karlsruhe Institute of Technology, 76128 Karlsruhe, Germany}}
	\affiliation{\mbox{Institute for Quantum Materials and Technologies, Karlsruhe Institute of Technology, 76021 Karlsruhe, Germany}}
	\affiliation{Landau Institute for Theoretical Physics, 119334 Moscow, Russia}
\date{\today}

\begin{abstract}
Dynamics of the imbalance in occupations on even and odd sites of a lattice serves as one of the key characteristics for identification of the many-body localization transition.  In this work, we investigate the long-time behaviour of the imbalance in disordered one- and two-dimensional many-body systems in the regime of diffusive or subdiffusive transport. We show that memory effects originating from a coupling between slow and fast modes lead to a power-law decay of the imbalance, with the exponent determined by the diffusive (or subdiffusive) transport law and the spatial dimensionality. Analytical results are supported by numerical simulations performed on a two-dimensional system in the regime of weak localization. 
\end{abstract}

\maketitle

\section{Introduction}
\label{sec:introduction}

In the absence of interaction, disordered systems exhibit Anderson localization if the disorder is strong enough, and even for weak disorder in low spatial dimensionality, with the transport being fully suppressed in the thermodynamic limit~\cite{paper:absence_of_diffusion,paper:localization_in_two_d,evers08}. 
This phenomenon has its counterpart in the physics of highly excited states (those with finite energy density) of interacting disordered many-body systems---many-body localization (MBL)~\cite{paper:mbl_gmp,paper:mbl_baa,mbl_3,mbl_4,mbl_5,mbl_6,mbl_7}. Specifically, when the interaction is turned on, there is a critical strength of disorder $W_c$ above which the system exhibits MBL. According to the current understanding, for a short-range interaction, $W_c$ is finite in the thermodynamic limit for one-dimensional (1D) systems and increases slowly with the system size for higher spatial dimensionality [in particular, in two-dimensional (2D) systems].  For $W < W_c$ the system is ergodic, and the transport is of diffusive or subdiffusive character \cite{znidaric_diffusion, Agarwal2015anomalous,BarLev2015absence,
bar_lev_ergodic_side,gopalakrishnan2020dynamics}. 

The long-time dynamics of the imbalance $I(t)$ is used as one of the key markers of the MBL transition, both in experiments and in computational studies~\cite{imbalance_experiment_mbl_1, imbalance_experiment_mbl_2,mbl_elmer,elmer_2d}. For this purpose, an initial state of a charge-density-wave type is set up,
with a strong imbalance between the occupation numbers of even and odd sites. After this, the evolution governed by the Hamiltonian of the system takes place, and the time dependence of the imbalance is monitored. In the MBL phase, the imbalance saturates for long times, $t \to \infty$, at a non-zero value, reflecting non-ergodicity of the system. 
On the other hand, in the delocalized phase the imbalance tends to zero at $t \to \infty$, since ergodic systems loose memory of their initial state in the long-time limit. The goal of this work is to investigate what is the law of the decay of the imbalance in delocalized systems. 

Naively, one could expect that the imbalance decay is of exponential character. Indeed, the long-time transport in the delocalized regime is of essentially classical character, with the Anderson localization suppressed by interaction-induced dephasing (or when the localization length is much larger than the system size, as, e.g., in 2D systems at relatively weak disorder). When the classical dynamics is described within the formalism of the Boltzmann equation, inhomogeneities with a large wave vector $q$ decay exponentially fast, with a short characteristic time. This is easy to understand physically: for such an inhomogeneity to disappear, each particle should travel only a small distance of the order of a few lattice spacings. This should be contrasted with the slow decay of diffusive modes with $q \to 0$ that  requires that particles travel a large distance $ \sim \pi/q$.

Remarkably, numerical studies show a power-law decay of the imbalance in the ergodic phase \cite{Luitz2016extended,mbl_elmer,elmer_2d,Weidinger2018self-consistent,Poepperl2021dynamics,sierant2021observe}, strikingly different from the exponential decay that would follow from the above argument based on the Boltzmann equation.  As we show in this paper, such a behavior of the imbalance is in fact a very general property of a diffusive or subdiffusive disordered system. The key point is that there exist memory effects that are discarded by the Boltzmann equation (which has a Markovian character): a particle is scattered off an impurity, then moves diffusively through the system, and finally returns to scatter on the same impurity. It has been known since long ago that such quasiclassical memory effects and associated long-time tails are of crucial importance for some of the transport properties of a disordered system~\cite{Ernst1971long,Ernst1984long}. In particular, they may lead to strong magnetoresistance~ \cite{Mirlin1999strong} and to a zero-frequency anomaly in the {\it ac} conductivity~\cite{zero_freq_anomaly}. 

As we show below, memory effects also generate a coupling between fast and slow modes (as pointed out in Ref.~\cite{mbl_elmer}), which leads to a power-law decay of modes with large wave vectors $q$ and thus of the imbalance. The exponent of this decay is controlled by the time dependence of the return probability, so that, for a diffusive system,  the imbalance decays as $I(t) \propto t^{-d/2}$, where $d$ is the spatial dimensionality.  For a subdiffusive system, with the effective diffusion constant depending on the wave vector $q$ according to $D(q) \propto q^\beta$, the decay law of the imbalance is modified according to  $I(t) \propto t^{-d/(2+\beta)}$. Our theory, which substantiates earlier proposals~\cite{Gopalakrishnan2016Griffiths,mbl_elmer,Poepperl2021dynamics} for the role of hydrodynamic long-time tails, thus provides a relation between the exponents characterizing the mean square displacement and the imbalance decay that was observed in numerical simulations~\cite{bar_lev_ergodic_side,Poepperl2021dynamics}.

To support our analytical results and to demonstrate that in a disordered system the mode coupling generically leads to the power-law decay of the imbalance specified above, we have performed numerical simulations of a non-interacting 2D system. The choice of the 2D (rather than 1D) geometry allows us to explore numerically a non-interacting system in the regime $l \ll L \ll \xi$ (where $L$ is the system size, $l$ is the mean free path, and $\xi$ is the localization length), in which the system is diffusive and the localization effects are of minor importance. The advantage of considering a non-interacting system is rather obvious: we access the exact long-time dynamics in a big system (up to $200 \times 200$ sites). The numerical results confirm the analytically predicted decay of the imbalance, $I(t) \propto t^{-\gamma_I}$, governed by the memory effects, with the exponent $\gamma_I$ being somewhat below unity due to weak multifractality. In view of the generality of the memory-effect mechanism, our results equally apply to interacting systems. 

The structure of the paper is as follows. In Sec.~\ref{sec:imbalance}, we define the imbalance $I(t)$ and derive a relation between the long-time asymptotics of the imbalance and the density response function. In Sec.~\ref{sec:diagrammatics}, a diagrammatic calculation of the long-time tail in the imbalance resulting from memory effects is performed. The analytical results are supported by numerical simulations presented in Sec.~\ref{sec:numerics}. Our findings are summarized in Sec.~\ref{sec:summary}.

\section{Imbalance and its relation to the density response function}
\label{sec:imbalance}

In this section, we define the imbalance and derive a relation between its tail at long times and the density response function. We consider first a 1D lattice; a generalization to 2D geometry (or a higher dimensionality) is straightforward and discussed in the end of the section.

We consider the time-dependent imbalance between the particle numbers \(N_\text{even}(t)\), \(N_\text{odd}(t)\) at even and odd lattice sites $j$ normalized to the total number of sites \(N\),
\begin{align}
	I(t) &= \frac{\langle  N_\text{even}(t) - N_\text{odd}(t) \rangle}{N}\\
	&= \frac{1}{N} \sum_{\text{sites }j} \langle n_j(t) \rangle  (-1)^j
	\label{eq:imbalance_discrete}.
\end{align}
Here, the angular brackets denote the average over the quantum many-body state. Since we deal with disordered systems, the average $\langle \ldots \rangle$ below also includes the disorder average. We define the density $n_j(t)$ and its continuum version $n(x,t)$, as well as the corresponding Fourier transform 
\begin{equation}
\tilde{n}(q, t)= \sum_j e^{-\ci qaj} n_j(t) = \int dx \, e^{-\ci qx} n(x,t) \,,
\end{equation}
where $a$ is the lattice spacing.
The imbalance then reads:
\begin{align}
    I(t) = \frac{1}{n_0V} \left \langle \tilde{n}\left(q = \frac{\pi}{a}, t\right) \right\rangle,
    \label{eq:imbalance-n} 
\end{align}
where $n_0 = V^{-1} \langle \tilde{n}(q=0) \rangle$ is the conserved density and $V = N a$ is the system volume. 

Experimentally and numerically, one explores the relaxation (or its absence) in the system by setting up a maximally imbalanced initial state at $t=0$ that is then time-evolved with the Hamiltonian $H$ of the system until long times $t$. In this paper, we are interested in the long-time behavior of the imbalance in the delocalized phase where the system evolves towards an equilibrium state with a uniform density distribution, $I(t) \to 0$ at $t \to \infty$. To understand the form of this asymptotic tail, we can thus equivalently start from a state 
with only a small imbalance (i.e., that is close to equilibrium). 

In this way, we can reformulate the problem under consideration in terms of a linear response near the equilibrium. Specifically, let us consider the system at $t \le 0$ as an equilibrium state of the Hamiltonian $H_0 - H'$, where 
\begin{align}
    H^\prime= \frac{I_0}{\nu} \,  \tilde{n}(q,t) \,.
    \label{eq:density_modulation}
\end{align}
Here, $q$ is the wave vector of the charge-density wave, $I_0=  (n_0V)^{-1}\langle \tilde{n}(q, t=0) \rangle$
is the initial value of the imbalance, and $\nu$ is the density of states. The term $-H^\prime$ in the Hamiltonian describes a periodic potential that yields the initial imbalance $I_0$.  Now, at time $t$ we perform a quench by removing the term $-H^\prime$, which is equivalent to adding a perturbation $H^\prime$ to the initial Hamiltonian. The system then starts relaxing towards the equilibrium state of the Hamiltonian $H_0$ with a uniform density. i.e., zero imbalance. 

Applying the Kubo formula~\cite{bruus_flensberg} to obtain the density response to the perturbation~\eqref{eq:density_modulation}, we obtain
\begin{equation}
	\langle \tilde{n}(q, t) \rangle = 
	 \langle \tilde{n}(q, 0) \rangle \left[ 1 + \frac{1}{\nu} \int_0^t \dd{t^\prime} \chi(q, t - t^\prime)\right],
\label{eq:Kubo}     
\end{equation}
where $\chi(q, t)$ is the retarded density-density correlation function (equivalently, density-density response function),	
\begin{equation}	
	\chi(q, t) = - \frac{\ci \theta(t)}{V} \langle \left[ \tilde{n}(q, t), \tilde{n}(-q, 0)\right]\rangle \,,
\end{equation}
with the Heaviside theta function $\theta(t)$.
Note that, in the Kubo formula, we are supposed to average over the equilibrium state of the initial Hamiltonian, which is given by $H_0 + H^\prime$. However, since the analysis is performed to linear order in the small perturbation $H^\prime$, we can discard $H^\prime$ here and average over the equilibrium state of $H_0$ towards which the system evolves. 

Equations~\eqref{eq:imbalance-n} and \eqref{eq:Kubo} establish the relation of the long-time tail of the imbalance with the density response function. 
An extension of this relation to higher-dimensional systems is straightforward. In particular, for a 2D square lattice one can consider the checkerboard imbalance corresponding to a charge density wave with the wave vector $\mathbf{q} = (\pi/a, \pi/a)$ or the columnar imbalance with the wave vector $\mathbf{q} = (\pi/a, 0)$. The formulas \eqref{eq:imbalance-n} and \eqref{eq:Kubo} remain valid with the replacement of $q$ by the corresponding 2D wave vector $\mathbf{q}$.
This relation is used below for the analytical study of the imbalance decay.

At the Markovian level, and at sufficiently small values of the wave vector, $q \ll l^{-1}$, the density response function is given, in the momentum-frequency representation, by the well-known diffusive formula
\begin{equation}
\tilde{\chi}(q, \omega) = - \nu \frac{Dq^2}{Dq^2-i\omega} \,,
\label{chi-diff}
\end{equation}
where \(D =  v^2\tau/d \) is the diffusion constant with the particle velocity \(v\), transport scattering time $\tau$, and spatial dimensionality $d$.
Upon Fourier  transformation to time space, it yields
\begin{equation}
 \chi(q,t) = - \nu Dq^2 \exp(-Dq^2 t) \,.
 \end{equation}
Substituting this into Eq.~\eqref{eq:Kubo}, we get 
\begin{align}
    \langle \tilde{n}(q, t) \rangle &= \langle \tilde{n}(q, t) \rangle \exp(-D q^2 t)\,,
    \label{eq:exponential_density_decay}
\end{align}
and thus, according to Eq.~\eqref{eq:imbalance-n}, the exponential decay of the imbalance,
\begin{equation}
I(t) = I_0 \exp(-t/t_q) \,,
\end{equation}
with $t_q = 1/Dq^2$. 

With increasing $q$, the decay time $t_q$ becomes shorter, reaching a very short value $t \sim \tau$ at the ultraviolet border of the diffusive range of wave vectors, $q \sim l^{-1}$. 
For further increasing wave vectors, $q > l^{-1}$, the exponential decay $\exp(-t/t_q)$ with a short time $t_q \sim \tau$ remains valid in the quasiclassical contribution to the imbalance in the Markovian approximation. This is obvious physically (the waves with a shorter wave lengths are expected to decay faster) and is easy to check by using an explicit form of the modified diffusion propagator in such ballistic range of wave vectors (see, e.g., Ref.~\cite{ballistic_diffuson_1d} for 1D systems and Ref.~\cite{ballistic_diffuson_2d} for 2D systems). 

At large $q> l^{-1}$, this decay factor actually describes the envelope of the oscillatory quasiclassical Markovian imbalance. In particular, in the 1D case one gets: $I(t)\propto \exp(-t/2\tau)\sin(qvt)$. 
It should be noted that, on top of this quasiclassical contribution to $I(t)$, there is a purely quantum (described by only retarded or only advanced Green's functions in the diagrammatic language, see below) one, which also decays exponentially in time. However, in contrast to the quasiclassical term, at the momentum $q=\pi/a$ at half filling (i.e., $q=2 k_\text{F}$, where \(k_{\rm F}\) is the Fermi momentum), this term yields a non-oscillatory contribution to $I(t)$.
Its decay rate is given by the maximum of \(1 / \tau\) and temperature \(T\) (in our case \(T \tau \gg 1\)). This is similar to the decay of Friedel oscillations or magnetooscillations, which are also suppressed by both disorder and thermal averaging. 

Thus, at the level of the Boltzmann equation (i.e., in the Markovian approximation), the imbalance decays exponentially, with a very short decay time. However, as we show in Sec.~\ref{sec:diagrammatics} by a diagrammatic analysis, there exists a contribution of memory effects, which is discarded by this approximation.  Calculating this contribution, we demonstrate that the actual decay of the imbalance is of power-law form and determine the corresponding exponent.

\section{Diagrammatic analysis}
\label{sec:diagrammatics}

To calculate the long-time tail in the density response function [and thus in the imbalance in view of the relations \eqref{eq:imbalance-n} and \eqref{eq:Kubo}], we use the conventional diagrammatic technique for disordered systems. The calculation bears analogy with that of the zero-frequency anomaly of the conductivity in Ref.~\cite{zero_freq_anomaly}. The starting point for the calculation is the formula~\cite{bruus_flensberg} for the density response function expressed in terms of exact retarded and advanced Green's functions, $\mathcal{G}^{\mathrm{R},\mathrm{A}}$, in a given realization of disorder:
\begin{align}
   & \tilde{\chi}(q, \omega) = -\int \frac{\dd\varepsilon}{2\pi \ci}\, 
    \int \frac{\dd^d p}{(2\pi)^d}\, n_{\rm F}(\varepsilon)
    \notag
    \\
    &\ \times
    \left\{\left[\mathcal{G}^\mathrm{R}(\mathbf{p}+\mathbf{q},\varepsilon)-\mathcal{G}^\mathrm{A}(\mathbf{p}+\mathbf{q},\varepsilon)\right]\mathcal{G}^\mathrm{A}(\mathbf{p},\varepsilon-\omega)\right.
    \notag\\
    &\  +
    \left.
    \mathcal{G}^\mathrm{R}(\mathbf{p}+\mathbf{q},\varepsilon+\omega)
    \left[\mathcal{G}^\mathrm{R}(\mathbf{p},\varepsilon)-\mathcal{G}^\mathrm{A}(\mathbf{p},\varepsilon)\right]
    \right\},
     \label{chi-G}
   \end{align}
where \(n_{\rm F}(\varepsilon)\) is the Fermi function.
This general expression is then averaged over disorder realizations.
We model disorder by a white-noise potential with the strength 
$\Gamma = (2 \pi \nu \tau)^{-1}$,
where  \(\tau\) is the elastic scattering time (equal to the transport time in this disorder model).

At this point, we have to select the diagrams from the disorder average that dominate the density-response function in the long-time limit. In the conventional case of a low external momentum $q$, the ladder sum of disorder lines features a pole and thus yields a diffuson, governing the long-time tail. The diffuson propagator describes the slow spread of the mean square displacement and is associated with the particle returns in arbitrary long times. At large external momentum, however, the situation is different: as pointed out in the previous section, the ladder sum in this case decays exponentially with time and thus does not describe a long-term memory.

The memory effects---that control the long-time tails that we are investigating---originate from the following type of processes. A particle is scattered by an impurity, then performs a diffusive motion during a long time $t$, which results in its return to the original position, where it is scattered again by the same impurity. By transferring the large external momentum via one or several impurity lines across the impurity ladder, the latter can again carry a small momentum, which results in a long-time tail.

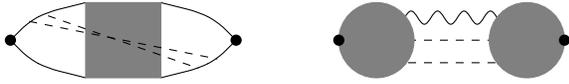
\begin{figure}[tb!]
    \begin{tikzpicture}
        \draw [rounded corners] (-1, 0.5) -- (-0.75, 0.25) -- (-0.5, 0.125) -- (0, 0);
        \draw [rounded corners] (-1, 0.5) -- (-0.75, 0.75) -- (-0.5, 0.875) -- (0, 1);
        \draw [rounded corners] (1, 0) -- (1.5, 0.125) -- (1.75, 0.25) -- (2, 0.5);
        \draw [rounded corners] (1, 1) -- (1.5, 0.875) -- (1.75, 0.75) -- (2, 0.5);
        \filldraw[gray] (0,0) rectangle (1, 1);
        \draw [dashed] (-0.75, 0.75) -- (1.75, 0.25);
        \draw [dashed] (-0.5, 0.825) -- (1.5, 0.125);
        \node at (-1, 0.5) [circle,fill,inner sep=1.5pt]{};
        \node at (2, 0.5) [circle,fill,inner sep=1.5pt]{};
    \end{tikzpicture}
    \hspace{1cm}
    \begin{tikzpicture}
        \draw [snake it] (-1, 0.3) -- (1, 0.3);
        \draw [dashed] (-1, 0) -- (1, 0);
        \draw [dashed] (-1, -0.3) -- (1, -0.3);
        \filldraw [gray] (-1, 0) circle (0.5);
        \filldraw [gray] (1, 0) circle (0.5);
        \node at (-1.5, 0) [circle,fill,inner sep=1.5pt]{};
        \node at (1.5, 0) [circle,fill,inner sep=1.5pt]{};
    \end{tikzpicture}
	\caption{{\it Left:} Example of a diagram contributing to the long-time tail of the density response function (and, thus, of the imbalance). The shaded box is the diffuson. It is crossed by one or several (two in the shown example) lines representing return events of a particle to the same scatterer(s) after moving diffusively for a long time $t$. 
	  {\it Right:} Another representation of the same diagram. The diffuson is shown here by a wavy line. }
	\label{fig:general_return_diagram}
\end{figure}

In a more general form, the scattering on a single impurity is replaced by scattering events on a few (two, three, \ldots) nearby impurities. An example of a corresponding diagram is shown in the left panel of Fig.~\ref{fig:general_return_diagram}. The shaded box in this diagram is the diffuson (the ladder built out of impurity lines). Two dashed lines crossing the diffuson correspond to a repeated scattering of the particle on two nearby impurities after completing a closed diffusive path. The same diagram is shown, in a different way, in the right panel of the same figure, with the diffuson represented by a wavy line.

Every additional crossing line adds an additional smallness of the order \(\mathcal{O}(1 / (k_{\rm F} l))\). Disorder ladders can only be added in combination with more crossing lines, since inserting one as a vertex correction would lead to exponential suppression of the diagram at high external momentum in the long-time limit. For this reason it suffices in the long-time and large mean-free path limits to calculate the sum of diagrams with the least number of disorder- and diffuson lines, which does not vanish.

Let us start by considering the lowest-order processes describing repeated scattering on a single impurity. They are represented by diagrams with a diffuson crossed by a single impurity line. 
For weak disorder, these diagrams yield the dominant contribution to the memory effects. For not so weak disorder, diagrams with two or three crossing impurity lines may give a comparable contribution but this will only correct the overall numerical prefactor, without affecting the result in any essential way.  

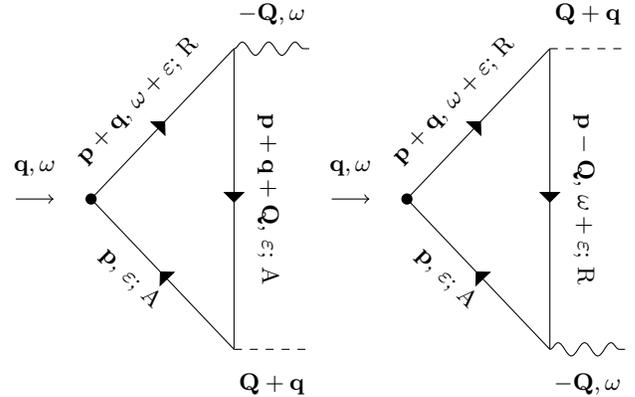
\begin{figure}[H]
	\centering
	 \begin{tikzpicture}
	 \coordinate (A) at (0, 0);
	 \coordinate (B) at (1.9, 2);
	 \coordinate (C) at (1.9, -2);
	 \begin{scope}[every node/.style={sloped,allow upside down}]
	    \draw[->] (-1, 0) -- (-0.5, 0) node [midway, fill=white, above=0.2cm] {\(\mathbf{q}, \omega\)};
	    \draw (A) -- node {\midarrow} (B) node [midway, fill=white, above=0.2cm] {\(\mathbf{p} + \mathbf{q},\, \omega + \varepsilon;\, {\rm R}\)};
	    \draw (B) -- node {\midarrow} (C) node [midway, fill=white, above=0.2cm] {\(\mathbf{p} + \mathbf{q} + \mathbf{Q},\, \varepsilon;\, {\rm A}\)};
	     \draw (C) -- node {\midarrow} (A) node [midway, fill=white, above=0.2cm, label={[rotate=-45, label distance=-0.5cm]\(\mathbf{p},\, \varepsilon;\, {\rm A}\)}] {};
	    \draw [snake it] (B) -- (2.9, 2) node [midway, fill=white, above=0.2cm] {\(-\mathbf{Q}, \omega\)};
	    \draw [dashed] (C) -- (2.9, -2) node [midway, fill=white, below=0.2cm] {\(\mathbf{Q} + \mathbf{q}\)};
	    \node at (A) [circle,fill,inner sep=1.5pt]{};
	\end{scope}
    \end{tikzpicture}
    \begin{tikzpicture}
    \coordinate (A) at (0, 0);
	 \coordinate (B) at (1.9, 2);
	 \coordinate (C) at (1.9, -2);
	 \begin{scope}[every node/.style={sloped,allow upside down}]
	    \draw[->] (-1, 0) -- (-0.5, 0) node [midway, fill=white, above=0.2cm] {\(\mathbf{q}, \omega\)};
	    \node at (A) [circle,fill,inner sep=1.5pt]{};
	    \draw (A) -- node {\midarrow} (B) node [midway, fill=white, above=0.2cm] {\(\mathbf{p} + \mathbf{q},\, \omega + \varepsilon;\, {\rm R}\)};
	    \draw (B) -- node {\midarrow} (C) node [midway, fill=white, above=0.2cm] {\(\mathbf{p} - \mathbf{Q},\, \omega + \varepsilon;\, {\rm R}\)};
	    \draw (C) -- node {\midarrow} (A) node [midway, fill=white, above=0.2cm, label={[rotate=-45, label distance=-0.5cm]\(\mathbf{p},\, \varepsilon;\, {\rm A}\)}] {};
	    \draw [dashed] (B) -- (2.9, 2) node [midway, fill=white, above=0.2cm] {\(\mathbf{Q} + \mathbf{q}\)};
	    \draw [snake it] (C) -- (2.9, -2) node [midway, fill=white, below=0.2cm] {\(-\mathbf{Q}, \omega\)};
	\end{scope}
    \end{tikzpicture}
	\caption{Triangle vertices \(V_1(\mathbf{q}, \mathbf{Q}, \varepsilon, \omega)\) and \(V_2(\mathbf{q}, \mathbf{Q}, \varepsilon, \omega)\) entering Eq.~\eqref{eq:W0}.
Here, \(\mathbf{q}\) is the external momentum and \(-\mathbf{Q}\) is the diffuson momentum. Disorder lines are dashed, diffuson lines are wavy. Retarded and advanced Green's functions are marked with R and A, respectively.
The external vertices of the density response function are marked by thick dots.}
\label{fig:LO_vertices}
\end{figure}

We analyze the density response function   $\tilde{\chi}\left(\mathbf{q}, \omega\right)$ at low frequencies (which correspond to long times $t$). 
 The sum of the diagrams with a diffuson and an impurity line inserted in all possible ways (corresponding to a rescattering on this impurity after executing the diffusive motion) can be written as 
\begin{align}
    \tilde{\chi}\left(\mathbf{q}, \omega\right) & 
    = - \ci\, \omega B(\mathbf{q}) \int \frac{\dd[d]{Q}}{(2\pi)^d} \Lambda^\text{diff}(\mathbf{Q}, \omega) \,,
    \label{eq:chi_lo}
\end{align}
where \(\Lambda^{\rm Diff}(\mathbf{Q}, \omega)\) is the diffuson,
\begin{align}
    \Lambda^{\rm Diff}(\mathbf{Q}, \omega) &= \frac{1}{2 \pi \nu \tau^2} \frac{1}{DQ^2 - \ci \omega} \,,
    \label{simple-diffuson}
 \end{align}
 and  the prefactor $B(\mathbf{q})$ is given by
\begin{align} 
\label{eq:Bq}
   B(\mathbf{q}) & =  \Gamma \int_{-\infty}^{\infty} \frac{\dd{\varepsilon}}{2\pi }  \left[ - \pdv{n_{\rm F}(\varepsilon)}{\varepsilon} \right] \, b (\mathbf{q}, \varepsilon)
 \end{align}
 with   
    \begin{align}    
   b (\mathbf{q}, \varepsilon) & = \lim_{\mathbf{Q} \rightarrow 0} \lim_{\omega \rightarrow 0} \left[V_1(\mathbf{q}, \mathbf{Q}, \varepsilon, \omega) + V_2(\mathbf{q}, \mathbf{Q}, \varepsilon, \omega)\right]^2.
    \label{eq:W0}
\end{align}
Here,    \(V_1\) and \(V_2\) are the vertex functions represented by the triangular diagrams  shown in Fig.~\ref{fig:LO_vertices}. In this Figure, \(\mathbf{q}\) is the external momentum and \(-\mathbf{Q}\) is the diffuson momentum, with the difference \(\mathbf{q} + \mathbf{Q}\) carried by the impurity line crossing the diffuson (as discussed in the introduction).
Since $-\mathbf{Q}$ and $\omega$ are the small momentum and frequency carried by the diffuson, we can discard them when calculating the vertices  \(V_1\) and \(V_2\), as indicated in Eq.~\eqref{eq:W0}. The formulas \eqref{eq:Bq} and \eqref{eq:W0} are obtained under the assumption that the vertex function $V_1+V_2$ has a finite limit at 
$\mathbf{Q} \rightarrow 0$ and  $\omega \rightarrow 0$. We show below by an explicit calculation that this is indeed generically the case. 

The vertex functions \(V_1(\mathbf{q}, \mathbf{Q}, \varepsilon, \omega)\) and \(V_2(\mathbf{q}, \mathbf{Q}, \varepsilon, \omega)\) entering Eq.~\eqref{eq:W0} are given by 
(see Fig.~\ref{fig:LO_vertices})
\begin{align}
    V_1(\mathbf{q}, \mathbf{Q}, \varepsilon, \omega) &= \int \frac{\dd[d]{p}}{(2\pi)^d} 
    G^{\rm R}(\mathbf{p} + \mathbf{q}, \omega + \varepsilon)  
    \nonumber \\
    & \times 
    G^{\rm A}(\mathbf{p} + \mathbf{q} + \mathbf{Q}, \varepsilon)\,
    G^{\rm A}(\mathbf{p}, \varepsilon)  \,,
     \label{eq:vertex_1}\\
    V_2(\mathbf{q}, \mathbf{Q}, \varepsilon, \omega) &=\int \frac{\dd[d]{p}}{(2\pi)^d} 
    G^{\rm R}(\mathbf{p} + \mathbf{q}, \omega + \varepsilon)
    \nonumber \\
    & \times 
    G^{\rm R}(\mathbf{p} - \mathbf{Q}, \omega + \varepsilon)\, 
    G^{\rm A}(\mathbf{p}, \varepsilon)    \,,
    \label{eq:vertex_2}
    \end{align}
    where \(G^{\rm R}\) and \(G^{\rm A}\) are, respectively, the disorder-averaged retarded and advanced Green's functions $\mathcal{G}^{\mathrm{R},\mathrm{A}}$ from Eq.~\eqref{chi-G}, 
 \begin{align}   
    G^{\rm R}(\mathbf{p}, \varepsilon) &= \frac{1}{\varepsilon - \xi(\mathbf{p}) + \frac{\ci }{2\tau}} = G^{\rm A *}(\mathbf{p}, \varepsilon) \,,
\end{align}
and \(\xi(\mathbf{p})\) is the dispersion relation of the clean system.

Substituting Eq.~\eqref{eq:Bq} into Eq.~\eqref{eq:chi_lo} and performing the Fourier transformation from the frequency to the time domain, we get
\begin{equation}
\chi(\mathbf{q}, t) = \frac{B(\mathbf{q})}{2\pi \nu \tau^2} \frac{\partial}{\partial t} P(t) \,,
\label{eq:chi-Pt}
\end{equation}
where $P(t)$ is the diffusive return probability,
\begin{align}
P(t) & =  \int \frac{\dd[d]{Q}}{(2\pi)^d}\int \frac{d\omega}{2\pi} e^{-i\omega t} \, \frac{1}{DQ^2 - i\omega} \nonumber \\ & = \int \frac{\dd[d]{Q}}{(2\pi)^d} \, e^{-DQ^2t} \,,
\end{align}
equal to
\begin{align}
P(t) & = (4\pi Dt)^{-d/2} \,.
\label{eq:Pt-normal}
\end{align}
By definition, $P(t)$ is the probability density for a diffusing particle that starts at a point $\mathbf{x_0}$ at time $t=0$ to be found at the same point after time $t$. Substituting  Eq.~\eqref{eq:chi-Pt}  into Eqs.~\eqref{eq:imbalance-n} and \eqref{eq:Kubo}, we get
  \begin{align}
\frac{I(t)}{I_0} = c P(t) \,, 
\label{eq:imbalance-Pt}
\end{align}   
where 
  \begin{align}
c = \frac{B(\mathbf{q})}{2\pi \nu^2 \tau^2} \,.
\label{eq:prefactor-imbalance}
\end{align}  
According to Eq.~\eqref{eq:imbalance-Pt}, the long-time tail of the imbalance is given (up to a coefficient) by the return probability $P(t)$. 

Let us recall at this point that our main motivation is the physics on the ergodic side of the MBL transition. There, the interaction generates dephasing, thus destroying the localization. For this reason, we discard localization effects in the above diagrammatic analysis. This is especially important in 1D geometry, where the diffusive regime does not exist in the absence of interaction since the localization length of a non-interacting system is of the order of the mean free path. In the case of higher-dimensional systems, $d \ge 2$, our analysis applies also to non-interacting systems at not too strong disorder, such that the system is delocalized, i.e., the localization length is much larger than the system size.

We have obtained the formulas \eqref{eq:chi-Pt} and \eqref{eq:imbalance-Pt} that relate the long-time tails in the density response function to the return probability:
$\chi(\mathbf{q}, t) \propto \partial P(t) / \partial t$ and $I(t) \propto P(t)$. 
While we have assumed conventional diffusive motion during the time $t$ described by a simple diffuson \eqref{simple-diffuson}, the effect is expected to remain valid in a more complex situation, when the particle executes a subdiffusion between the original scattering and the return to the same impurity. We will thus use these relations below in such, more general sense. 

\subsection{1D systems}

We evaluate now the general formulas for the density response function and the imbalance for the case of a 1D system. To simplify the calculation,
 it is convenient to linearize the dispersion relation 
 \begin{align}
     \xi(p) \simeq \begin{cases}
        \xi_+(p) = (p - k_{\rm F})\,v, & p > 0 \,,\\
        \xi_-(p) = -(p + k_{\rm F})\,v, & p < 0 \,,
     \end{cases}
     \label{linearization}
 \end{align}
 where the branches $\xi_+$ and $\xi_-$ correspond to right-moving and left-moving particles. The linearization does not affect the result in any essential way (up to an overall numerical prefactor of order unity).  Upon linearization, we can easily carry out the integrations in Eqs.~\eqref{eq:vertex_1} and \eqref{eq:vertex_2}. We recall that we are interested in the limit \(\omega \rightarrow 0\), \(Q \rightarrow 0\). Further, we set the external momentum to be \(q = \pi / a\). For this value of $q$, the particle always switches the branch at the external vertex. We denote the triangle vertices with \(- \rightarrow +\) change of the branch at the external vertex (going along the arrow in Fig.~\ref{fig:LO_vertices}, i.e., from $G^\text{A}$ to $G^\text{R}$) by \(V_1^\mp\), \(V_2^\mp\) and those with the change \(+ \rightarrow -\) by \(V_1^\pm\) and \(V_2^\pm\). The calculation outlined in Appendix~\ref{sec:vertices_1d_imbalance} yields
    \begin{align}
     V_1^\mp(\varepsilon) + V_2^\mp(\varepsilon) 
        & = \frac{2\tau}{v}\, \frac{2k_{\rm F} v+ 2\varepsilon - \pi v/ a }
        { (2k_{\rm F} v+ 2\varepsilon - \pi v/ a)^2+1 / \tau^2}  \,, 
        \label{V1mp+V2mp}
     \end{align}
and, similarly,      
    \begin{align}     
        V_1^\pm(\varepsilon) + V_2^\pm(\varepsilon) &= -\frac{2\tau}{v}\,\frac{2k_{\rm F} v - 2\varepsilon - \pi v / a}{(2k_{\rm F} v - 2\varepsilon - \pi v / a)^2+1 / \tau^2}.
        \label{V1pm+V2pm}
    \end{align}

For definiteness, we assume the half filling, \(k_{\rm F} = \pi / 2a \), in the following. (For a different density, the result remains the same, up to a prefactor.)
We note in passing that, for half filling, the vertices \eqref{V1mp+V2mp} and \eqref{V1pm+V2pm} vanish exactly at $\varepsilon=0$ (which is a manifestation of an extra symmetry related to Umklapp scattering),
but are finite for any finite energy. Therefore, at nonzero temperatures, the vertex factor given by Eq.~\eqref{eq:W0} is nonzero.  
Combining the contributions of the \(- \rightarrow +\) and \(+ \rightarrow -\) processes to the triangle vertices, we get
for the prefactor $B(q = \pi/a)$ in Eq.~\eqref{eq:chi_lo}
\begin{align}
    B =  \Gamma \left(\frac{8 \tau}{v}\right)^2 \int_{-\infty}^\infty \frac{\dd{\varepsilon}}{2\pi} \left[-\pdv{n_{\rm F}(\varepsilon)}{\varepsilon}\right]\, \left(\frac{\varepsilon}{1 / \tau^2 + 4 \varepsilon^2} \right)^2.
\label{eq:1D-B}
\end{align}

Since the initial density-wave state is highly excited, it corresponds to a high temperature $T$, comparable to the band width $J$. We thus make an assumption $T \tau \gg 1$ to calculate the prefactor. The integral in Eq.~\eqref{eq:1D-B} is then easily calculated, yielding $B = \tau^2 / 2 \pi v T$. 
This gives for the prefactor in Eq.~\eqref{eq:chi-Pt}
\begin{equation}
\frac{B}{2\pi \nu \tau^2} = \frac{1}{4\pi T} \,,
\end{equation}
and thus $ c = v/4T$ for the prefactor $c$ in Eq.~\eqref{eq:prefactor-imbalance}. This calculation of the prefactor (involving linearization of the spectrum) is controllable for $T \ll J$. For an estimate, we can, however, put here $T \sim J$, which yields $c \sim a$. 

The above analysis, leading to the power-law decay of the imbalance, 
\begin{equation}
 I(t)    \propto P(t) \propto t^{-1/2},
\end{equation}
applies to the diffusive regime of transport that takes place at a sufficiently weak disorder (well below the MBL transition) in interacting disordered systems \cite{znidaric_diffusion}. At the same time, numerical studies show that a major part of the ergodic phase of such systems is characterized by subdiffusive transport \cite{znidaric_diffusion, Agarwal2015anomalous,BarLev2015absence,
bar_lev_ergodic_side,gopalakrishnan2020dynamics}.
In this paper, we do not analyze a microscopic mechanism leading to subdiffusion in a particular model. Instead, we assume that the subdiffusive behavior holds and model it on a phenomenological level by introducing a modified diffusion propagator:
\begin{align}
	\Lambda^\text{diff}(Q, \omega) &\to \Lambda^\text{subdiff}_\beta(Q, \omega) \sim \frac{1}{\nu\tau^2} \frac{1}{D(Q) Q^2 - \ci \omega} \,,
	\label{eq:anomalous_diffuson}\\
	D(Q) &=  \tilde{D} |Q|^\beta \,. \label{eq:anomalous_diffuson_2}
\end{align}
Here \(\beta > 0\) is the exponent controlling the subdiffusive character of the transport: $\beta =0$ corresponds to normal diffusion, while $\beta \gg 1$ corresponds to the very slow transport as found near the MBL transition.  The propagator \eqref{eq:anomalous_diffuson} corresponds to the fractional diffusion equation \cite{levi_flights_main_reference}; the associated 
 mean square displacement 
 \begin{align}
    r_2(t) = \left\langle \int \dd{x} x^2 \bar{n}(x, t) \right\rangle
\end{align}
 reads (see, e.g., Ref.~\cite{levi_flights_main_reference}):
\begin{align}
   r_2(t) \sim 	{(\tilde{D}t)^{ \frac{2}{2 + \beta}}}.
   \label{eq:r2-1D-subdiffusion}
\end{align}

Now, we analyze the long-time tail in the imbalance. As found above, it is proportional to the return probability $P(t)$ in the case of conventional diffusion. We argue that this result still holds true for subdiffusion. Indeed, this is expected because diffusive and subdiffusive processes are established at long times (long spatial scales), while the vertex functions at high external momentum $q\sim k_{\rm F}$ are determined by large momenta, i.e., by short time (or spatial) scales. Therefore, microscopic details of the diffusive or subdiffusive process can plausibly be assumed to be irrelevant for the vertices. Using the anomalous-diffusion propagator \eqref{eq:anomalous_diffuson}, we get for the return probability
\begin{equation}
P(t) \sim {(\tilde{D}t)^{ - \frac{1}{2 +\beta}}}.
\label{eq:Pt-1D-anom}
\end{equation}
Substituting this into Eqs.~\eqref{eq:chi-Pt} and \eqref{eq:imbalance-Pt}, we obtain the asymptotics of the density response function,
\begin{equation}
\chi(q, t) = \frac{B(q)}{2\pi \nu \tau^2} \frac{\partial}{\partial t} P(t) \propto t^{-1 - \frac{1}{2 +\beta} } \,,
\label{eq:chi-1D-anom}
\end{equation}
and of the imbalance,
  \begin{align}
\frac{I(t)}{I_0} = c P(t) \propto t^{- \frac{1}{2 +\beta} } \,.
\label{eq:imbalance-1D-anom}
\end{align}  

The slow power-law decay of the imbalance \eqref{eq:imbalance-1D-anom} is in agreement with numerical findings on the ergodic side of the MBL transition \cite{Luitz2016extended,mbl_elmer,elmer_2d,Weidinger2018self-consistent,Poepperl2021dynamics,sierant2021observe}.  Comparing    Eq.~\eqref{eq:r2-1D-subdiffusion} and Eq.~\eqref{eq:imbalance-1D-anom}, we see a relation between the exponent $\gamma_x$ characterizing the mean square displacement, $r_2(t) \propto t^{\gamma_x}$, and the exponent $\gamma_I$ describing the imbalance decay, $I(t) \propto t^{-\gamma_I}$.  Specifically, we obtain
\(\gamma_x = 2 / (2 + \beta)\)  and  \( \gamma_I = 1 / (2 + \beta)\),  with the ratio  \(\gamma_I / \gamma_x = 1 / 2\), independent of the subdiffusive exponent \(\beta\). This exponent relation was proposed in Ref.~\cite{bar_lev_ergodic_side} and is in reasonable agreement with numerical results on long-time dynamics in large systems obtained within the time-dependent Hartree-Fock approximation in Ref.~\cite{Poepperl2021dynamics}. 

It should be emphasized, however, that the above derivation of the relation between the exponents is based on the assumption that the anomalous diffusion coefficient $D(q)$ in Eq.~\eqref{eq:anomalous_diffuson} depends on the momentum $q$ and not on frequency $\omega$.   This leads to Eq.~\eqref{eq:Pt-1D-anom} for the return probability and, thus, to the scaling \eqref{eq:imbalance-1D-anom} of the imbalance. A more complex situation, with the anomalous diffusion constant $D(q,\omega)$ showing (at small $\omega$ and relatively large $q$) a scaling with both $q$ and $\omega$, corresponds to multifractality. In such a situation (that it is characteristic, in particular, to Anderson-transition critical points) the scaling of the return probability $P(t)$ is characterized by an exponent that is not directly determined by the exponent of the mean square displacement. We will return to this issue below. 

\subsection{2D systems}
\label{sec:analytics-2D}

We extend now the analysis to 2D systems, $d=2$. One natural extension of the imbalance to 2D systems on a square lattice is the checkerboard-imbalance 
\begin{align}
    I_{\rm check}(t) &= \sum_{i, j} (-1)^{i + j} \frac{\langle n_{(i, j)}(t) \rangle}{N} \,.
    \label{eq:imbalance-2D-check}
\end{align}
Here \(i\) and \(j\) enumerate the rows and columns of the system, respectively. Taking the continuum limit in analogy to the 1D case, we find, in analogy with 
Eq.~\eqref{eq:imbalance-n},
\begin{align}
     I_{\rm check}(t) = \frac{1}{n_0V} \left \langle \tilde{n}\left(q_x = \frac{\pi}{a}, \, q_y = \frac{\pi}{a}, \,  t\right) \right\rangle.
    \label{eq:imbalance-n-2D-check}
\end{align}
Alternatively, one can consider the columnar imbalance~\cite{elmer_2d}    $I_{\rm col}(t)$ corresponding to the density wave with wave vector \(q_x = \pi / a\) and \(q_y = 0\). 
Our analytical treatment applies equally to both    $I_{\rm check}(t)$ and $I_{\rm col}(t)$, so we use below the notation $I(t)$ to refer to any of them. For numerical calculations, we indicate which of the imbalances is shown. 

Equations \eqref{eq:chi-Pt} and  \eqref{eq:imbalance-Pt} give the tails of the density-response function and of the imbalance in terms of the return probability $P(t)$. In the case of normal diffusion, the return probability is given by Eq.~\eqref{eq:Pt-normal}.  This yields the scaling 
$$\chi(\mathbf{q},t) \propto t^{-2}$$ 
for the density-response function and 
\begin{equation}
I(t) \propto t^{-1}
\end{equation}
for the imbalance. Estimating the coefficients, we get $B/ 2\pi\nu\tau^2 \sim 1/J$ for the coefficient in Eq.~\eqref{eq:chi-Pt} and $c \sim a^2$ for the coefficient in Eq.~\eqref{eq:imbalance-Pt}. 

For a subdiffusive transport modelled by the anomalous
diffusion propagator, Eqs.~\eqref{eq:anomalous_diffuson} and \eqref{eq:anomalous_diffuson_2}, we obtain the results analogous to Eqs.~\eqref{eq:Pt-1D-anom},  \eqref{eq:chi-1D-anom}, and  \eqref{eq:imbalance-1D-anom}, with a replacement of the exponent $1/(2 +\beta)$ by  $2/(2 +\beta)$.  
For the ratio of the exponents, this yields $\gamma_I / \gamma_x = 1$. Clearly, a similar consideration in arbitrary spatial dimensionality would give 
$$\gamma_I / \gamma_x = d/2.$$ 

As was already pointed out in Sec.~\ref{sec:introduction}, the 2D geometry allows us to consider a regime of (nearly) diffusive transport also in the absence of interaction. Indeed, even though the non-interacting system gets localized in the thermodynamic limit, the localization length $\xi$ is much larger than the mean free path $l$ when the disorder is sufficiently weak. The transport in the regime $l \ll L \ll \xi$ has then diffusive character (with weak-localization corrections for which the system size \(L\) serves as an infrared cutoff \cite{lee_ramakrishnan}), and the decay of imbalance can be investigated within the non-interacting picture. This problem is studied numerically below in Sec.~\ref{sec:numerics}. The non-interacting character of the model allows us to consider rather large system sizes ($200 \times 200$) within exact diagonalization. We focus on times \(t\) much smaller than the time of diffusive spreading through the system. Before turning our attention to the numerical simulations, let us discuss the implications of the weak localization for the above analytical results.

The weak localization leads to a frequency-dependent logarithmic correction to the diffusion constant:
\begin{equation}
D(\omega) \simeq D_0 \left( 1 - \frac{1}{\pi k_F l} \ln \frac{1}{\omega \tau} \right) \,.
\label{eq:D-WL}
\end{equation}
Note that the asymptotics of the mean square deviation $r_2(t)$ is controlled by the diffusion constant $D(q,\omega)$ at small $\omega$ and small $q$, with $Dq^2 \sim \omega$, so that we can put $q=0$ in Eq.~\eqref{eq:D-WL}. In the regime of frequencies where the correction is relatively small, we can rewrite Eq.~\eqref{eq:D-WL} as
\begin{equation}
D(\omega) \simeq D_0 (\omega\tau)^{\frac{1}{2\pi g}} \,,
\end{equation}
where we introduced the dimensionless conductance $g = k_F l /2$. This implies for the mean square deviation 
\begin{equation}
r_2(t) \sim t^{1 - \frac{1}{2\pi g}} \,,
\end{equation}
i.e., a weak-localization correction to the exponent: $\gamma_x = 1 - 1/2\pi g$. 

The tail of the return probability $P(t)$ is controlled by weak multifractality of 2D systems (which is responsible for the behavior of the diffusion constant $D(q,\omega)$ at small $\omega$ and relatively large $q$). The corresponding multifractal exponent is~\cite{wegner80,falko95,mirlin00} $d_2 = 2 - 2/\pi g$,
yielding 
\begin{equation}
P(t) \sim t^{-d_2 /d} = t^{-1 + \frac{1}{\pi g}} \,,
\end{equation}
and thus $\gamma_I = 1 - 1 / \pi g$.   We see that the corrections to $\gamma_x$ and $\gamma_I$ are different (by factor of 2), and thus the exponents $\gamma_x$ and $\gamma_I$  deviate not only from unity but also from each other.

\section{Numerical results}
\label{sec:numerics}

As discussed above, the numerics in this paper is restricted to non-interacting 2D systems. We calculate the long-time asymptotics of both the checkerboard imbalance and the columnar imbalance starting from the corresponding maximum-imbalance states. In addition, we calculate the linear-response  density response function \(\chi(q_x, q_y, t)\),  verifying thereby the relation~\eqref{eq:imbalance-Pt}  between the long-time tail of the imbalance and density response function. This also allows us to check that the power-law tail of the density response function has the same form for all momenta $\mathbf{q}$.

We consider a square lattice of \(N = L \times L\) sites described by the Hamiltonian 
\begin{align}
    H &= J \sum_{\mathbf{r}, \mathbf{r'}} \delta_{\langle \mathbf{r}, \mathbf{r'} \rangle} c_\mathbf{r}^\dagger c_\mathbf{r'} + \sum_{\mathbf{r}} \varepsilon_\mathbf{r} c_\mathbf{r}^\dagger c_\mathbf{r} \,,\label{eq:numerics_hamilton}
 \end{align}
where  $\mathbf{r}$ and $\mathbf{r'}$ label sites of the square lattice and
\begin{align}    
    \delta_{\langle \mathbf{r}, \mathbf{r'} \rangle} &= \begin{cases}
        1\,, & \mathbf{r}, \mathbf{r'} \text{ nearest neighbors},\\
        0\,, & \text{else}.
    \end{cases}
\end{align}
We set \(J = a= 1\). The onsite potential values  \(\varepsilon_\mathbf{r}\) are uncorrelated random numbers drawn from a random uniform distribution in the interval \([-W, W]\).

We analyze the numerical results based on the predictions for the density response function  and the imbalance  at long times,
\begin{align}
    \chi(q, t) &= \chi_0 \exp(-t / t_q) + \frac{\chi_1}{t^{1+\gamma_I}} \,, \label{eq:fit_function_density_density}\\
    I(t) &= I_0 \exp(-t / t_q) + \frac{I_1}{t^{\gamma_I}}\,. \label{eq:fit_function_imbalance}
\end{align}
The first terms in these formulas correspond to the exponentially decaying contribution from the Markovian approximation. Here, we keep these terms in addition to long-time tails, in order to be able to describe the case of sufficiently small values of $q$, such that the exponential decay is not yet strong at times addressed by numerical simulations.
The second terms in Eqs. \eqref{eq:fit_function_density_density} and \eqref{eq:fit_function_imbalance} are the long-time asymptotics governed by return processes. The exponent $\gamma_I$ is slightly below unity, $\gamma_I = 1- 1/\pi g$, as discussed in Sec.~\ref{sec:analytics-2D}. 

Since we are interested in the diffusive regime, we first need to identify an appropriate disorder strength. If the disorder is too weak, a density perturbation would spread ballistically; on the other hand, too strong disorder would lead to strong localization for considered system sizes. To identify the diffusive regime, we calculate the mean square displacement
\begin{align}
    r_2(t) &= \left\langle \sum_{j = 1}^L \sum_{i=1}^L R_{i,j} \left[n_{(i, j)}(t) - n_{(i, j)}(t=0)\right] \right\rangle\,, \label{eq:disk_msq}\\
    R_{i, j} & = [(i - i_0)^2 + (j - j_0)^2].
\end{align}
Here, \(n_{(i, j)}(t)\) is the particle density at site \((i, j)\) at time \(t\), with \(i\) and \(j\) labeling rows and columns, respectively, and angular braces denote an average over disorder configurations. The site \((i_0, j_0)\) is the original position of the density packet. Specifically, we initialize the system with %
\begin{align}
    n_{(i, j)}(t=0) &= \delta_{i,i_0} \delta_{j,j_0} \,.
    \label{eq:initial}
\end{align}
In order to minimize finite-size effects, we choose the site \((i_0, j_0)\)  to be located in the center of the system.

\begin{figure}[tb]
    \centering
    \includegraphics[width=0.95\columnwidth]{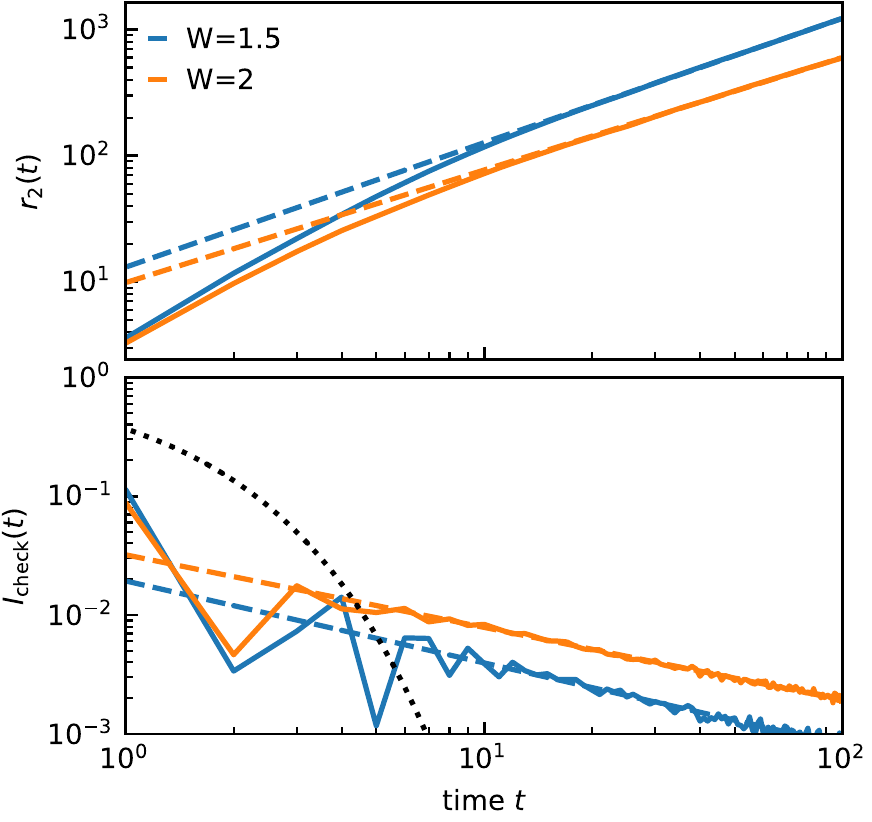}
    \caption{Mean square displacement \(r_2(t)\) (upper panel) and checkerboard imbalance \(I_{\rm check}(t)\) (lower panel) as functions of time for 2D systems with disorder strengths $W=1.5$ and $W=2$. Calculations were performed on a square lattice of \(201 \times 201\) sites with open boundary conditions; averaging over 5 disorder configurations was done. The dashed lines in the upper panel are power-law fits, $r_2(t) \sim  t^{\gamma_x}$, yielding  $ \gamma_x = 0.88$ for \(W = 2\) and $ \gamma_x = 0.98$ for \(W = 1.5\). The dashed lines in the lower panels are power-law fits $I_{\rm check}(t) \sim t^{-\gamma_I}$, yielding $\gamma_I = 0.61$ for $W=2$ and $\gamma_I = 0.69$ for $W=1.5$.  The black dotted line shows an exponential decay \(\exp(-t/\tau)\) with $\tau =1$ in units of the hopping time for comparison.}
    \label{fig:2d_msq_and_I_check}
\end{figure}

The results for \(r_2(t)\) for disorder strengths $W=1.5$ and $W=2$ are presented in the upper panel of Fig.~\ref{fig:2d_msq_and_I_check}: We find the asymptotic power-laws \(r_2(t) \sim t^{0.88}\) for \(W = 2\) and \(r_2(t) \sim t^{0.98}\) for \(W=1.5\).  The exponents are slightly below unity, in agreement with the expectation $\gamma_x = 1 - 1/2\pi g$. Therefore, these values of disorder correspond to the diffusive regime with weak-localization corrections. For stronger disorder ($W=2$), the correction is more significant as expected. 
Using \(r_2(t) = 4D_0t\) and \(D_0 = v^2 \tau / 2\) at time $t \approx 10$ at which the diffusion is fully established, we get an estimate for the mean free time:
$\tau \approx 2$ for $W=1.5$ and $\tau\approx 1$ for $W=2$. The mean free time decreases with increasing $W$ approximately as $1/W^2$, as expected for relatively weak disorder.  We have also verified that if the initial state is chosen as a 1D domain wall and the corresponding 1D mean square displacement is calculated, the same results are obtained as for the disk mean square displacement~\eqref{eq:disk_msq}.

The following comment is in order here. Since our initial condition contains single-particle states with different energies,
our numerical procedure effectively involves the corresponding averaging. The dominant contribution comes from the broad central part of the band, where the dimensionless conductance $g$ weakly depends on energy and where the majority of states is located. At the same time, one expects also a contribution of band tails, where $g$ is smaller, so that the states have a localization length shorter than our system size. For the mean square displacement $r_2(t)$ this would only induce a small correction to the effective diffusion constant. At the same time, the contribution of localized states should lead to a saturation of the imbalance at long times, $t\to \infty$. Thus, by inspecting the behavior of the imbalance, one can numerically find out whether the localized states from the band tails are essential for the dynamics on a given time scale. We will see below that, within the time range of our numerics, $ t = 10^2$, the role of band tails is negligible, even for our stronger disorder, $W=2$. Therefore, within this time range, we essentially probe the physics associated with the majority of states in the central part of the band. This justifies our description, Eqs.~\eqref{eq:fit_function_density_density} and \eqref{eq:fit_function_imbalance}. Indeed, we will see below that the predicted power laws for the imbalance and density response function are nicely observed in numerical simulations.

%

%

\begin{figure}[t!]
    \centering
    \includegraphics[width=0.95\columnwidth]{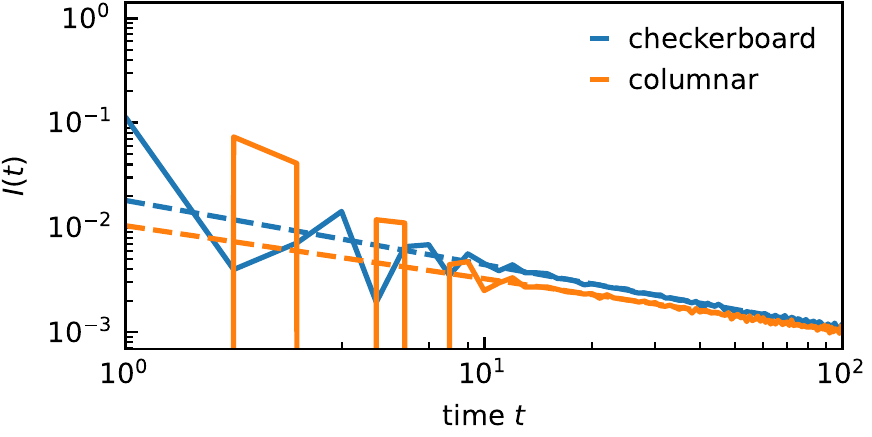}
    \caption{Checkerboard and columnar imbalance as functions of time at disorder \(W = 1.5\). Simulations were performed on square lattices of \(101 \times 101\) and \(100 \times 100\) sites, respectively, with open boundary conditions and with averaging over 60 disorder configurations. The dashed lines are power-law fits 
$I(t) \sim t^{-\gamma_I}$, yielding $\gamma_I = 0.61$ for the checkerboard imbalance and $\gamma_I = 0.51 $ for the columnar imbalance. 
    }
    \label{fig:comparison_check_stripe}
\end{figure}

\subsection{Imbalance}
\label{sec:num-imbalance}
After having identified the diffusive regime by inspecting the mean square displacement, we turn to the numerical analysis of the imbalance.
The checkerboard imbalance for \(W = 1.5\) and \(W=2\) is shown in the lower panel of Fig.~\ref{fig:2d_msq_and_I_check}. A power-law decay of the imbalance is clearly observed.
Fitting the imbalance tail to a power-law $\propto t^{-\gamma_I}$, we find  $\gamma_I \approx 0.69$ for $W=1.5$ and $\gamma_I \approx 0.61$ for $W=2$.  The values of the exponent $\gamma_I$ are somewhat below unity, in agreement with the analytical prediction $\gamma_I = 1 - 1/\pi g$.
The deviation of $\gamma_I$ from unity is larger for larger disorder, as expected. Further, the deviations of $\gamma_I$ from unity are larger than the respective deviations of $\gamma_x$, again in agreement with the analytical expectations. 

As pointed out above, the imbalance does not exhibit any saturation within the considered time window (even though it drops down to a relatively small value $\sim 10^{-3}$). This shows that strongly localized states in the band tails do not play any essential role in this time range. In Appendix~\ref{Appendix-C}, we explicitly check this statement by evaluating the fraction of strongly-localized states contributing to the imbalance dynamics in the transient time window $t \le 100$. We also demonstrate there that the conductance in the band of extended states only slightly deviates from the value in the band center. As a result, the contributions of different energies to the imbalance produce, in our transient time window, a function that is indistinguishable from a simple power law. 

In order to emphasize the significance of the slow, power-law decay, we also show the Markovian result \(\exp(-t / \tau)\) with \(\tau = 1\) in the plot (black dotted line). 
On the scale of $ t \approx 10$, this exponential contribution becomes negligible ($\sim 10^{-4}$). For our largest times, $t \approx 100$, it drops down to a value as small as $\sim 10^{-40}$. Our numerical results therefore clearly confirm an important role of classical memory effects in the imbalance of a disordered system. Furthermore, the predicted difference between the imbalance- and mean-square displacement exponents, \(\gamma_I\) and \(\gamma_x\), is observed numerically.

As shown in Fig.~\ref{fig:comparison_check_stripe}, the behavior of the columnar imbalance is very similar to that of the checkerboard imbalance. Indeed, they are very close numerically and show almost the same power-law decay, with $\gamma_I \approx 0.51$ for the checkerboard imbalance and $\gamma_I \approx 0.61$ for the columnar imbalance. The system size in this figure is \(N = 101 \times 101\), i.e., smaller than in Fig.~\ref{fig:2d_msq_and_I_check} (where $N = 201 \times 201$). A slightly smaller value of $\gamma_I $ for the checkerboard imbalance in comparison with Fig.~\ref{fig:2d_msq_and_I_check} is thus attributed to finite-size effects.

\subsection{Density response function}
We have also performed numerical simulations of the density response function \(\chi(q_x, q_y, t)\), which is predicted to decay at long times as $t^{-1-\gamma_I}$,  see Eq.~\eqref{eq:fit_function_density_density}. Note that this prediction applies for any value of the momentum $(q_x,q_y)$. To make a direct connection with the numerical analysis of the imbalance
in Sec.~\ref{sec:num-imbalance},
we carry out a linear-response calculation with respect to a thermal state with the chemical potential chosen in the center of the band, $\mu=0$, and with a temperature of the order of the band width; see Appendix~\ref{appendix:density_response} for details.

Instead of directly investigating the long-time tail of the correlator \(\chi(q_x, q_y, t)\), we perform its numerical integration to obtain the long-time behaviour of the imbalance at the considered wave vector [see Eqs.~\eqref{eq:imbalance-n}, \eqref{eq:Kubo}]:
\begin{align}
   I_{q_x, q_y}(t) &\propto  \int_0^t \dd{t'} \chi(q_x, q_y, t')- \tilde{\chi}_0 \,.
   \label{eq:integrated-dd-response}
\end{align}
Here the constant $\tilde{\chi}_0$ is equal to the zero-frequency limit of the density response function, $\tilde{\chi}_0 \equiv \tilde{\chi}(\omega=0, q_x, q_y)$, which ensures 
$I_{q_x, q_y}(t) \to 0$ at $t \to \infty$. 
 In order to characterize the long-time tails, we fit the integrated density response
 $\int_0^t \dd{t'} \chi(q_x, q_y, t')$
 in a late-time window \(t \in [20, 100]\) to the function
\begin{align}
    f(t) = f_0 + f_1 t^{-\gamma_{I}}
    \label{ft}
\end{align}
with fitting parameters \(f_0\), \(f_1\), and \(\gamma_I\). The constant \(f_0\) corresponds to 
$\tilde{\chi}_0$ of Eq.~\eqref{eq:integrated-dd-response}
and is subtracted to get the imbalance. 
 In this way, we obtain the imbalance \(I_{q_x, q_y}(t)\) and the imbalance exponent \(\gamma_I\) for the whole range of momenta $(q_x,q_y)$.
 
 %
 %
\begin{figure}[ht!]
    \centering
    \includegraphics[width=\linewidth]{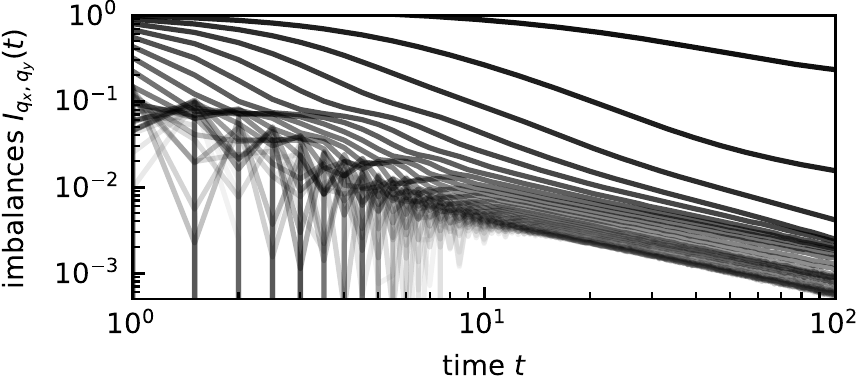}
    \includegraphics[width=\linewidth]{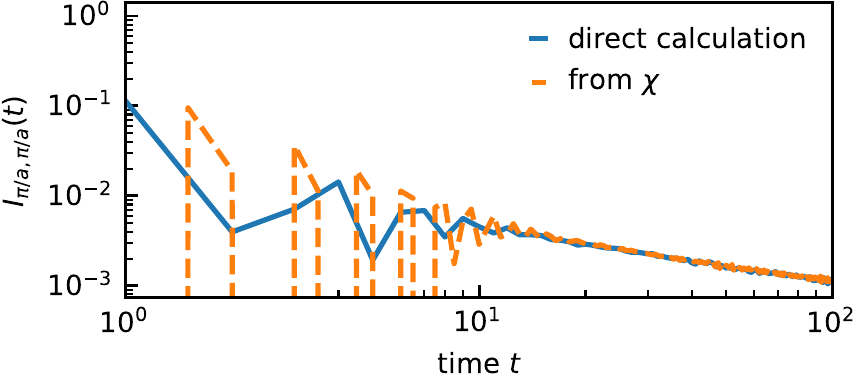}
        \caption{Upper panel: Numerical results for the imbalance \(I_q(t)\equiv I_{q_x = q, q_y = q}(t)\) obtained according to
        Eq.~\eqref{eq:integrated-dd-response}
        from the density response function    
        for \(W = 1.5\), \(L = 80\), and temperature \(T = 3\), with periodic boundary conditions. The data was averaged over about \(500\) disorder realizations.  Imbalance \(I_q(t)\) is calculated for momenta \(q=q_n= 2\pi n/L\) with \(n \in [2, 39]\). The numerical factor between imbalance and integrated density response function is obtained from a comparison of the long-time tail of the largest-$q$ curve to the directly calculated imbalance result (cf.  Fig.~\ref{fig:comparison_check_stripe}). Lower panel: Comparison between the checkerboard imbalance directly calculated from the time evolution of a checkerboard state (with  \(L =101\)) and the checkerboard imbalance from the density response function. For this comparison, the shift constant and factor were determined by fitting the integrated density response function to a power-law with the same exponent as found for the checkerboard imbalance from the direct calculation.} 
    \label{fig:chi_fits}
\end{figure}


In Fig.~\ref{fig:chi_fits}, we show \(I_q(t) \equiv I_{q_x=q,q_y=q}(t)\) at temperature \(T = 3\) for a square system with $L = 80$ and disorder $W=1.5$, for momenta \(q=q_n=2 \pi n / L\) with $n = 2, 3, \ldots, 39$. For this plot the integrated density was rescaled by a factor determined from comparison of the large-\(q\) tails to the directly calculated imbalance. (Since the actual factor between imbalance and integrated response depends on the momentum, this can lead to the small-\(q\) curves exceeding unity at short times.) The values of momenta increase from top to bottom. For the lowest momenta, the power-law decay can barely be observed within the time window of the simulation, since the exponential contribution decays slowly.  For larger momenta, the exponential contribution decays very quickly, so that \(I_q( t)\) is governed by the power-law tail starting already from rather short times. We observe that, for sufficiently large \(q\), all imbalance curves become parallel straight lines in the long-time limit, confirming the momentum independence of the exponent. 

\begin{figure*}[t!]
      \includegraphics[width=1.75\columnwidth]{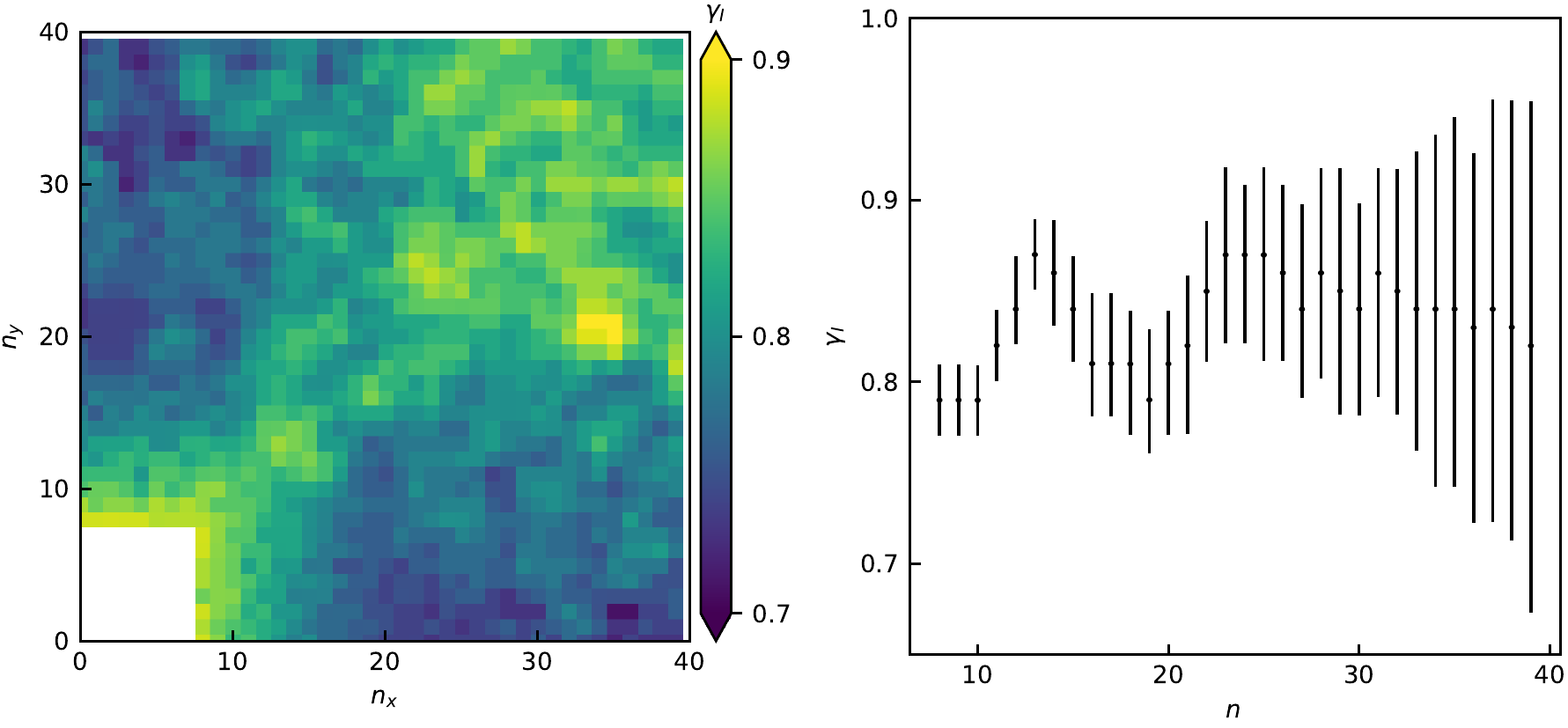}
    \caption{Left panel: Tomography of power-law exponents $\gamma_I(q_x,q_y)$ for imbalance obtained according to
    Eq.~\eqref{eq:integrated-dd-response}
    from the density response function.  
    Exponents are obtained from fitting the long-time tail to a power-law in the time interval \(t \in [20, 100]\). The range $(n_x < 8, n_y < 8)$ is excluded since the time window of the simulation does not allow us to extract reliably the power-law exponents at these momenta.  Right panel: Exponents 
    $\gamma_I(q_x = q_y = q)$ describing the power-law tails of the imbalance curves shown in Fig.\ref{fig:chi_fits}.
    The error bars are the intervals for which the mean square error of the corresponding fit is up to five percent larger than the mean square error of the optimal fit.} 
    \label{fig:chi_fits_tomo}
\end{figure*}

In Fig.~\ref{fig:chi_fits_tomo}, we show the values of the exponent $\gamma_I(n_x, n_y)$ corresponding to \(I_{q_{n_x}, q_{n_y}}(t)\). The left panel shows a color map of the exponent as a function of \(n_x\) and \(n_y\) for \(n_x, n_y = 2,3, \ldots, 39\).
The right panel displays the diagonal exponents, \(n_x = n_y = n\). As expected from the imbalance plots, the time window of our simulation does not suffice to find reliably the power-law exponents in the low momentum sector \(n \lesssim 8\), as the exponential component decays too slowly. This region is therefore excluded in both panels. The error bars in the right panel show the range of exponents, for which the mean square error of the obtained fit deviates by up to five percent from the optimum fit.
The numerical results are consistent with the analytical predictions that $\gamma_I$ is independent of $q$ and is somewhat below 1 (because of weak multifractality). Further, the numerical value \(\gamma_I(q_{39}) \approx 0.8\), corresponding to the checkerboard imbalance, is in a good agreement with \(\gamma_I \approx 0.7\) extracted from the direct checkerboard imbalance calculation
in Sec.~\ref{sec:num-imbalance}. This agreement is also demonstrated in the lower panel of Fig.~\ref{fig:chi_fits} where the checkerboard imbalance obtained by direct simulations and from the density response function are compared.
For this plot, the shift constant \(\tilde{\chi}_0\) was obtained by fitting the integrated density response to a power-law with the exponent found for the directly calculated imbalance (slightly differing from the optimal-fit exponent for the integrated density); the overall scaling factor was fixed by comparing the tails.

\section{Summary and conclusions}
\label{sec:summary}

In this paper, we have shown that memory effects lead to a power-law asymptotic tail of the imbalance in disordered systems, $I(t) \propto t^{-\gamma_I}$. 
We have derived a relation between the imbalance and the density response function and showed that $I(t) \propto P(t)$, where $P(t)$ is the probability for a particle to be found at the original point after a long time $t$  (``return probability'').  In the case of normal diffusive transport, the analysis based on classical memory effects related to diffusive returns yields $\gamma_I = d/2$, where $d$ is the spatial dimensionality. Having in mind the ergodic side of the MBL transition, we have also considered the regime of subdiffusive transport. Specifically, we used its phenomenological modelling in terms of a momentum-dependent diffusion constant, $D(q) \propto q^\beta$, which yields the mean square deviation $r_2(t) \propto t^{\gamma_x}$ with \(\gamma_x = 2 / (2 + \beta)\).  For the imbalance in this situation, we obtained a power-law decay with the exponent
 \( \gamma_I = d / (2 + \beta)\),  implying the ratio  \(\gamma_I / \gamma_x = d / 2\), independent of the subdiffusive exponent \(\beta\). 
 
 To complement the analytical results, we have performed numerical simulations for non-interacting disordered 2D systems. In these simulations, we have chosen a disorder range that ensures the diffusive character of transport for considered system sizes. More accurately, in view of the weak-localization effects, the transport is ``weakly subdiffusive'', i.e., the exponent $\gamma_x$ is slightly below unity. For such systems, we have demonstrated a crucial role of memory effects in the long-time behavior of the imbalance and found a power-law decay of the imbalance. The corresponding exponent $\gamma_I$ shows a downward deviation from unity, which is related to weak multifractality of eigenstates of 2D non-interacting disordered systems. This deviation leads to a weak violation of the relation $\gamma_I = \gamma_x$.
 
Our results explain the slow, power-law decay of the imbalance on the ergodic side of the MBL transition, as observed in numerical simulations of 1D disordered interacting systems
 \cite{Luitz2016extended,mbl_elmer,elmer_2d,Weidinger2018self-consistent,Poepperl2021dynamics,sierant2021observe}.  The relation $\gamma_I = \gamma_x$ that we find by modelling the subdiffusive transport by a diffusion constant $D(q) \propto q^\beta$ is consistent with numerical observations \cite{bar_lev_ergodic_side,Poepperl2021dynamics}.
 The subdiffusive transport in this class of system is usually attributed to Griffiths effects related to rare strongly localized spots. Our analysis is, however, rather general and shows that, whatever the mechanism of the subdiffusion is, it will lead to the corresponding slow decay of the imbalance due to mode coupling induced by the memory effects. 
 
 A slow decay of the imbalance was also numerically observed on the ergodic side of the MBL transition in 2D systems ~\cite{elmer_2d,Poepperl2021dynamics}. In this case, it was found that the corresponding effective exponent $\gamma_I$ increases with time, saturating at the value $\gamma_I = 1$ at long times.
This is consistent with the relation $\gamma_I = \gamma_x$,  since in 2D geometry the Griffiths effects cannot suppress the conventional diffusion ($\gamma_x=1$). An increase of $\gamma_I$ towards unity at intermediate times is a transient effect attributed to trapping of particles at rare localized spots~\cite{Gopalakrishnan2016Griffiths,Poepperl2021dynamics}.
  
A slow, power-law decay of the imbalance was numerically found also for 1D quasiperiodic systems. Specifically, it was observed \cite{Poepperl2021dynamics} that the exponent $\gamma_I$ increases with time, saturating at the value $\gamma_I = 1$.  This is in consistency with the relation $\gamma_I = (d/2)\gamma_x$, in view of the ballistic character of transport ($\gamma_x = 2$) in quasiperiodic systems. It is worth pointing out, however, that our analysis in this paper was performed for truly random systems, so that its application to quasiperiodic systems should be viewed as a conjecture. Further work in this direction is needed, especially in view of the importance of quasiperiodic systems for experimental investigations.

A weak violation of the relation $\gamma_I = (d/2)\gamma_x$ in 2D non-interacting disordered systems in the weak-localization regime poses the question as to whether the relation is exact on the ergodic side of the MBL transition. The mechanism related to quantum coherence of single-particle states, which is responsible for multifractality in 2D non-interacting systems, should not be relevant for the ergodic interacting systems at high temperature, in view of decoherence. This provides an expectation that the relation $\gamma_I = (d/2)\gamma_x$ strictly holds (for the exponents characterizing the limiting long-time behavior) in the ergodic phase of an interacting disordered system.  In fact, Ref.~\cite{Gopalakrishnan2016Griffiths} identified other power-law contributions related to trapping of particles by localized spots in 1D geometry. These contributions are, however, subleading (i.e., decaying faster) in comparison with that studied in the present paper, and thus do not affect our derivation of the relation $\gamma_I = (d/2)\gamma_x$. Further computational and experimental work towards a systematic verification of the relation between the exponents $\gamma_I$ and $\gamma_x$ on the ergodic side of the MBL transition would be of much interest. 

\begin{acknowledgements}
We are grateful to E.~V.~H. Doggen, J.~F. Karcher, D.~G. Polyakov, and K.~S. Tikhonov for discussions. This work was supported by Deutsche Forschungsgemeinschaft (DFG) via grant No. GO 1405/6-1.
\end{acknowledgements}

\onecolumngrid

\appendix

\section{Calculation of vertices in the non-Markovian term in imbalance in 1D systems}
\label{sec:vertices_1d_imbalance}

In this Appendix, we calculate the sum of triangle vertices $V_1+V_2$, Fig.~\ref{fig:LO_vertices},
for 1D systems with linearized dispersion (\ref{linearization}). For the vertices $V_{1,2}^\mp$  that switch the branch from $-$ to $+$ at external momentum $q=\pi/a$, we have the following integral over the momentum in infinite limits, $-\infty<p<\infty$: 
    \begin{align}
     V_1^\mp(\varepsilon) &+ V_2^\mp(\varepsilon) =
         \int_{-\infty}^\infty \frac{\dd{p}}{2\pi} \,
         \frac{1}{\varepsilon- \xi_-(p)  - \frac{\mathrm{i}}{2\tau}} \, \frac{1}{\varepsilon - \xi_+(p+\pi/a) + \frac{\mathrm{i}}{2\tau}}  
          \left[\frac{1}{ \varepsilon -\xi_-(p) + \frac{\mathrm{i}}{2\tau}} + \frac{1}{\varepsilon - \xi_+(p+\pi/a) - \frac{\mathrm{i}}{2\tau}}\right]
         \nonumber \\
        & =
        \int_{-\infty}^\infty \frac{\dd{p}}{2\pi} \, \frac{1}{\varepsilon + (p + k_{\rm F})v  - \frac{\mathrm{i}}{2\tau}}\,  \frac{1}{\varepsilon - (p + \pi / a - k_{\rm F})v + \frac{\mathrm{i}}{2\tau}}  
        \left[\frac{1}{ \varepsilon + (p + k_{\rm F})v + \frac{\mathrm{i}}{2\tau}} + \frac{1}{\varepsilon - (p + \pi / a - k_{\rm F})v - \frac{\mathrm{i}}{2\tau}}\right].
       \label{V1mp+V2mp-int} 
       \end{align}
Each term contains the poles in the upper and lower half-planes. The contour integration yields
\begin{align}
     V_1^\mp(\varepsilon) &+ V_2^\mp(\varepsilon) =
     \frac{\tau}{v}\left(\frac{1}{2\varepsilon+2k_F v-\pi v/a + \mathrm{i}/\tau}+\frac{1}{2\varepsilon+2k_F v-\pi v/a - \mathrm{i}/\tau}\right),
\end{align}
which results in Eq.~(\ref{V1mp+V2mp}) of the main text.
Equation (\ref{V1pm+V2pm}) is obtained analogously. 

Let us now explicitly demonstrate the vanishing of the sum of vertices $V_1^+(\varepsilon) + V_2^+(\varepsilon)$ that do not switch the branch $+$ to branch $-$.
The calculation is analogous to the above:
  \begin{align}
     V_1^+(\varepsilon) &+ V_2^+(\varepsilon) =
         \int_{-\infty}^\infty \frac{\dd{p}}{2\pi} \,
         \frac{1}{\varepsilon- \xi_+(p)  - \frac{\mathrm{i}}{2\tau}} \, \frac{1}{\varepsilon - \xi_+(p+\pi/a) + \frac{\mathrm{i}}{2\tau}}  
          \left[\frac{1}{ \varepsilon -\xi_+(p) + \frac{\mathrm{i}}{2\tau}} + \frac{1}{\varepsilon - \xi_+(p+\pi/a) - \frac{\mathrm{i}}{2\tau}}\right]
         \nonumber \\
        & =
        \int_{-\infty}^\infty \frac{\dd{p}}{2\pi} \, \frac{1}{\varepsilon - (p - k_{\rm F})v  - \frac{\mathrm{i}}{2\tau}}\,  \frac{1}{\varepsilon - (p + \pi / a - k_{\rm F})v + \frac{\mathrm{i}}{2\tau}}  
        \left[\frac{1}{ \varepsilon - (p - k_{\rm F})v + \frac{\mathrm{i}}{2\tau}} + \frac{1}{\varepsilon - (p + \pi / a - k_{\rm F})v - \frac{\mathrm{i}}{2\tau}}\right].
       \label{V1p+V2p-int} 
       \end{align}
Again, each of the two terms taken separately has poles in the upper and lower half-planes. However, the sum of the terms vanishes exactly after the contour integration:
\begin{align}
     V_1^+(\varepsilon) &+ V_2^+(\varepsilon) =
     \frac{\mathrm{i}}{v}\left(\frac{1}{-\pi v/a + \mathrm{i}/\tau}\,\frac{1}{\mathrm{i}/\tau}
     +\frac{1}{\pi v/a - \mathrm{i}/\tau}\,\frac{1}{\mathrm{i}/\tau}\right)=0.
     \label{zero-vertex}
\end{align}
Clearly, the same cancellation also occurs for the branch $\xi_-$ of left-movers.

\section{Additional numerical checks to Sec.~\ref{sec:numerics}: Fraction of localized states and energy dependence of the conductance in calculations of the imbalance}
\label{Appendix-C}

In Sec.~\ref{sec:numerics}, we numerically investigate memory effects in the (transient) diffusive regime of a 2D Anderson lattice. Our numerical results on the decay of the imbalance $I(t)$ in this regime are in agreement with the analytical prediction (Sec.~\ref{sec:analytics-2D}) of the power law behavior \(I(t) \propto t^{-\gamma_I}\), with exponent \(\gamma_I = 1 - 1 / (\pi g)\) for a non-interacting 2D system. Here, \(g\) is the conductance and the term \(-1 / (\pi g)\) in the exponent $\gamma_I$ originates from a weak-localization correction to the classical memory effects (\(\gamma_I = 1\)).

It might come as a surprise that the imbalance numerics presented in Sec.~\ref{sec:numerics} is described so well by a power law \(I(t) \propto t^{-\gamma_I}\), for the following two reasons. First, even though we consider not too strong disorder, there is a fraction of localized states, with localization lengths smaller than the size of the system. Some of these localized states (in the tails of the band) are so strongly localized that their localization length is already probed on the time scales of our numerical simulations. Since the imbalanced initial condition for the numerics (for example, a checkerboard pattern in the density) encompasses the full range of energies, such strongly-localized states would also contribute to the imbalance. Their  contribution is different from the  power law that is characteristic for delocalized states: a strongly localized state is expected to give a time-independent contribution. Second, as the conductance $g(\varepsilon)$ is generically energy-dependent, the initial condition for the imbalance implies averaging of the corresponding power-law decay over energy [here \(\nu(\varepsilon)\) is the density of states]:
\begin{align}
    I(t) \propto \int \dd{\varepsilon} \nu(\varepsilon)\, t^{-\left(1 - \frac{1}{\pi g(\varepsilon)}\right)}. \label{eq:av_powerlaw}
\end{align}

\begin{figure}[tb]
    \centering
    \includegraphics[width=0.95\columnwidth]{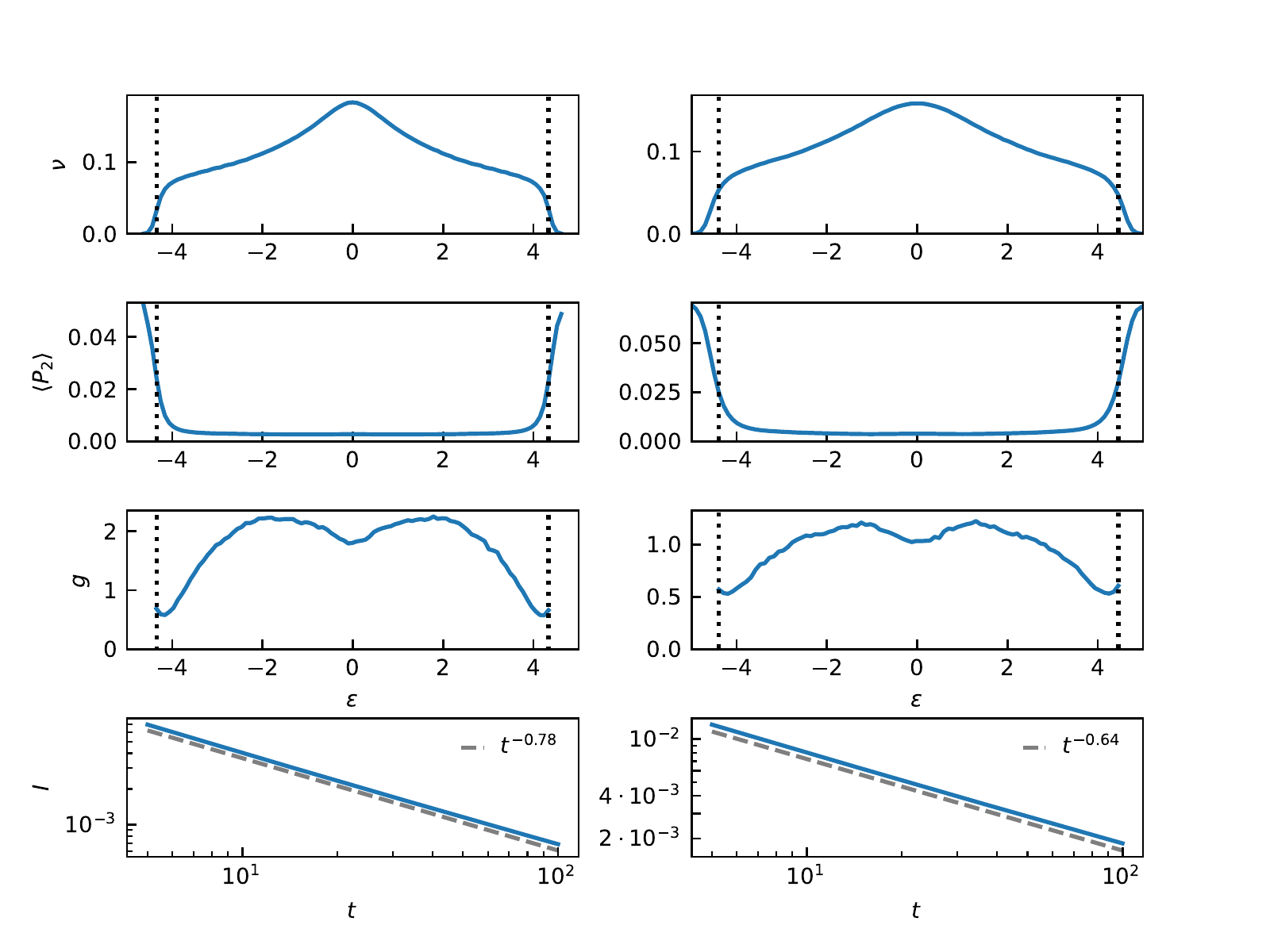}
    \caption{Density of states \(\nu\) (first row), inverse participation ratio \(P_2\) (second row), and conductance \(g\) (third row) as functions of energy $\varepsilon$; imbalance $I$ as a function of time on a double-logarithmic scale (fourth row).  These results are obtained from exact diagonalization after averaging over 4000 disorder realizations of Hamiltonian \eqref{eq:numerics_hamilton} in a system of \(40 \times 40\) sites at disorder \(W = 1.5\) (left column) and \(W = 2\) (right column). The vertical dotted lines separate the effectively delocalized states ($\xi \gg 10$) from the localized states ($\xi \lesssim 10$) in band tails.
    The imbalance was obtained (up to a constant prefactor) from \(\nu(\varepsilon)\) and \(g(\varepsilon)\) using Eq. \eqref{eq:av_powerlaw}. For convenience of comparison, the prefactors in the imbalance plots are chosen by fixing the values at time \(t = 10\) to the direct results in Fig. \ref{fig:2d_msq_and_I_check}. Dashed lines are power-law fits slightly shifted with respect to the imbalance curves (\ref{eq:av_powerlaw}) (solid) to make them easier to distinguish. }
    \label{fig:ipr_results}
\end{figure}

In Sec.~\ref{sec:numerics}, we provide explanation on why the numerical data for the imbalance in the transient diffusive time window are described so well by a power law 
with a single power law exponent \(\gamma_I = 1 - 1 / (\pi g)\). First, the fraction of strongly localized states in band tails is very small, so that they do not play any essential role in the considered time regime. The dominant contribution to the imbalance in this time range comes from the broad central region of the band (encompassing nearly all states), while the expected asymptotic saturation (due to strong localization) will set in at still longer times. Second, in this broad central part of the band, \(g\) is sufficiently large and depends only weakly on the energy, so that 
the average \eqref{eq:av_powerlaw}
is numerically almost indistinguishable from a simple power law. The purpose of this appendix is to demonstrate these two statements explicitly by numerically evaluating $g(\varepsilon)$ and  \(\nu(\varepsilon)\) across the energy band.

To this end, we calculate the eigenstates and eigenvectors of Hamiltonian \(\eqref{eq:numerics_hamilton}\) for \(4000\) disorder realizations with \(W = 1.5\) and \(W = 2\), in a system of \(N = 40 \times 40\) sites. From the eigenenergies we obtain the density of states \(\nu(\varepsilon)\), and for each eigenstate \(\psi(\varepsilon)\) at energy \(\varepsilon\) we determine the inverse participation ratio (IPR)
\begin{align}
    P_2(\varepsilon) = \sum_{i = 1}^N \left| \psi_i(\varepsilon) \right|^4.
\end{align}
Calculating the mean value $\langle P_2(\varepsilon) \rangle^2$ and the variance 
$ \text{var}[P_2(\varepsilon)]$ of $P_2(\varepsilon)$ (with respect to averaging over disorder realizations), we obtain information on the conductance $g(\varepsilon)$ and the localization length $\xi(\varepsilon)$, as we are now going to explain. For delocalized states (localization length much larger than the system size), the IPR is given by the random-matrix-theory value
\(\langle P_2 \rangle \approx 3 / N\), with a weak-localization correction. On the other hand, for strongly localized states (with \(\xi \ll  L\)), the IPR becomes much larger than this value. We can get an estimate of the localization length \(\xi\) of such a strongly localized state by assuming (for \( 1 \ll \xi \ll  L\)) that it spreads within the area \(\xi^2\), resulting in \(P_2 \sim 3 / \xi^2\). This allows us to estimate the contribution of the localized states to the conductance.
Further, we use IPR fluctuations to extract the conductance for the weakly localized states via \cite{mirlin00}
\begin{align}
    g(\varepsilon) = \sqrt{\mathcal{A}\frac{ \langle P_2(\varepsilon) \rangle^2}
    {\text{var}[P_2(\varepsilon)]}},
\end{align}
where \(\mathcal{A}\) is a numerical factor that depends on the spatial dimensionality and boundary conditions; in our case \(\mathcal{A} \approx 0.123\).   Using the obtained conductance and density of states, we numerically verify that energy averaging  \eqref{eq:av_powerlaw} indeed does not lead to any essential deviations from a simple power law (in the considered time window). 

Before presenting our numerical data, we point out that the localization length $\xi(\varepsilon)$ in the same 2D model was determined numerically by the transfer-matrix approach in Ref.~\cite{2d_conductance_numerics} (see upper panel of Fig. 2 there). The disorder used in Ref.~\cite{2d_conductance_numerics} was $W=2.5$ in our units, i.e., somewhat stronger than in our simulations. The results of Ref.~\cite{2d_conductance_numerics} show that, even for this stronger disorder, the fraction of strongly localized states with $\xi < 10$ (see below for the reason of the choice of this boundary) is very small. Furthermore, the conductance that can be estimated (from the one-loop formula) as $g \approx \pi^{-1} \ln \xi$ varies in a relatively narrow interval only, $1.3 \lesssim g \lesssim 1.7$, in the energy range $ |\varepsilon| < 3.5 $ comprising an overwhelming majority of all states. These results fully support the above two statements [formulated in the paragraph below Eq.~\eqref{eq:av_powerlaw}], in consistency with our numerics discussed below.

 In Fig.~\ref{fig:ipr_results}, we show the numerically obtained density of states (first row), average IPR (second row), conductance (third row), and imbalance decay obtained from Eq.~\eqref{eq:av_powerlaw} (fourth row) for \(W = 1.5\) (left column) and \(W = 2\) (right column). Inspecting the density of states, we observe that nearly all states lie within the energy band of the clean system, $|\varepsilon|<4$. 
Already from this figure, one sees that almost the whole band is effectively delocalized, with only a small fraction of strongly localized states in the tails.
From the IPR values, we find that states within \(|\varepsilon| \lesssim 4\) are ``delocalized'' from the finite-size perspective of the system, with \(\xi \gg L\) and thus \(\langle P_2 \rangle \sim 3 / N \sim 0.002\). 

Diffusion with \(D \sim 1\) over times $t\sim 100$ implies spreading over \(\sim \sqrt{100}= 10\) sites in each direction. Therefore, states with  \(\xi \gtrsim 10\) still appear delocalized in the time window explored with our numerics in Sec.~\ref{sec:numerics}. Placing a cut-off at \(\langle P_2 \rangle = 0.02 \lesssim 3 / 10^2\) on the density of states (dotted lines) to separate the strongly localized states, we find that the fraction of strongly localized states is indeed very small:
\(\approx 99\%\) of all states at \(W = 1.5\) and \(\approx 97\%\) of the states at \(W = 2\) are delocalized according to this criterion.  Further, for the conductance within the energy window corresponding to delocalized states, we find values between approximately \(0.6\) and \(2.3\) (\(0.5\) and \(1.2\)) for \(W = 1.5\) (\(W = 2\)). Note that the fact that $g(\varepsilon)$ has a local minimum at the band center is in full agreement with the results of Ref.~\cite{2d_conductance_numerics}.

Using the obtained results for $\nu(\varepsilon)$ and $g(\varepsilon)$, we numerically calculate the  energy-averaged imbalance curves according to
Eq.~\eqref{eq:av_powerlaw}, which are shown in the fourth row in Fig.~\ref{fig:ipr_results}.
We find that the resulting curves for both values of disorder are virtually indistinguishable from power laws  (dashed lines, slightly shifted for ease of comparing), with \(\gamma_I = 0.78\) at \(W = 1.5\) and \(\gamma_I = 0.65\) at \(W = 2\). These results are in good agreement with the values extracted from the direct imbalance simulations, \(\gamma_I = 0.69\) for \(W = 1.5\) and \(\gamma_I = 0.61\) for \(W = 2\), see  Fig.~\ref{fig:2d_msq_and_I_check}.

The fact that, despite the energy averaging \eqref{eq:av_powerlaw}, the imbalance is described so well by a single-power law is fully consistent with the observation that, in most of the band, the conductance $g(\varepsilon)$ varies only weakly around its band-center value $g(0)$ (see the third row in Fig.~\ref{fig:ipr_results}).
Specifically, we find that for $\sim 85\%$ of states, the conductance $g(\varepsilon)$ is within $\approx 25\%$ from its band-center value $g(0)$.

The localized states are expected to give a time-independent contribution $\sim 1/\xi^2$ to the imbalance. Even for our stronger disorder, we thus get an estimated contribution on the level of $10^{-4}$. This fully supports our interpretation of the numerics, as provided in Sec.~\ref{sec:numerics}. The power laws observed there are transient and will eventually saturate. However, the level at which saturation appears is very small ($\sim 10^{-4}$) and is not relevant in the considered time range (where the imbalance drops down only to $\sim 10^{-3}$).

\section{Numerical calculation of the density response function}
\label{appendix:density_response}

To calculate \(\chi(q, t)\) in a 2D non-interacting system numerically, we start from the definition in the site space:
\begin{align}
    \hat{\chi}_{\mathbf{r}, {\mathbf{r}^\prime}}(t) = - \mathrm{i} \theta(t) \langle [\hat{n}_\mathbf{r}(t), \hat{n}_{\mathbf{r}^\prime}(t)] \rangle.
    \label{eq:d_d_corr_numerics}
\end{align}
Here \(\mathbf{r}\) and \({\mathbf{r}^\prime}\) label the sites on the two dimensional grid and \(\hat{n}_{\mathbf{r}}(t)\) is the number operator in site space, with
\begin{align}
    \tilde{n}(q_x, q_y, t) &= \sum_{\mathbf{r}} e^{-\ci \mathbf{q} \mathbf{r} a} \hat{n}_{\mathbf{r}}(t).
\end{align}
Applying Wick's theorem, we find
\begin{align}
    \hat{\chi}_{\mathbf{r}, {\mathbf{r}^\prime}}(t) &
    = -2 \mathrm{Im} \left\{ \langle c^\dagger_{\mathbf{r}^\prime}(t) c_\mathbf{r} \rangle \langle c_{\mathbf{r}^\prime}(t) c_\mathbf{r}^\dagger \rangle \right\} = -2 \mathrm{Im}\left\{ [G^<_{{\mathbf{r}^\prime}, \mathbf{r}}(t, 0)]^* G^>_{{\mathbf{r}^\prime}, \mathbf{r}}(t, 0)\right\},
\end{align}
where we have identified the lesser and greater Green's functions \(G^<\) and \(G^>\). These Green's functions are time-evolved according to
\begin{align}
    G_{\mathbf{r}, {\mathbf{r}^\prime}}^\lessgtr(t, 0) &= \sum_{\mathbf{r}''}
    U_{\mathbf{r}, {\mathbf{r}^{\prime\prime}}}(t)\, G^\lessgtr_{{\mathbf{r}^{\prime\prime}},  {\mathbf{r}^\prime}}(0, 0),\\
    U_{\mathbf{r}, {\mathbf{r}^\prime}}(t) &= \left[\exp(-\ci Ht)\right]_{\mathbf{r}, {\mathbf{r}^\prime}},
\end{align}
where \(H\) is the Hamiltonian in the site space. 

We specify the initial condition in the eigenbasis of \(H\) (denoted with Greek indices), according to the Fermi distribution:
\begin{align}
    G^{\lessgtr}_{\mathbf{r}, {\mathbf{r}^\prime}}(0, 0) &= v_{\mathbf{r}, \alpha} G^{H, \lessgtr}_{\alpha, \beta} (v^\dagger)_{\beta, {\mathbf{r}^\prime}},\\
    G^{H, <}_{\alpha, \beta}(0, 0) &= \ci \frac{\delta_{\alpha, \beta}}{\exp[\beta(\varepsilon_\alpha - \mu)] + 1},
    \quad
    G^{H, >}_{\alpha, \beta}(0, 0) = \delta_{\alpha, \beta} \left[-\mathrm{i} + G^{H, <}_{\alpha, \beta}(0, 0)\right].
\end{align}

Here, 
\(\{\varepsilon_\alpha\}\) and \(\{v_{\mathbf{r}, \alpha}\}\) are the eigenenergies and eigenvectors of \(H\). 
The chemical potential \(\mu\) is chosen in the middle of the band, and the temperature \(T = 1/\beta\) is of the order of the bandwidth. We obtain \(\chi(q_x, q_y, t)\) by calculating the Fourier transform 
of \(\hat{\chi}_{\mathbf{r}-{\mathbf{r}^\prime}}(t)\) and performing the disorder average.

From \(\chi(q_x, q_y, t)\), the imbalance tails are extracted by using the relations
\begin{align}
    I_{\rm check}(t) &= \frac{1}{n_0 V}\langle \tilde{n}(q_x = \pi / a, q_y = \pi / a, t)\rangle, \qquad n_0 = \frac{1}{V} \langle \tilde{n}(q=0)\rangle,\\
	\langle \tilde{n}(q_x, q_y, t) \rangle &= 
	 \langle \tilde{n}(q_x, q_y, 0) \rangle \left[ 1 + \frac{1}{\nu} \int_0^t \dd{t^\prime} \chi(q_x, q_y, t - t^\prime)\right],
\end{align}
see Sec.~\ref{sec:imbalance}. 
We then perform numerical integration of \(\chi(q_x, q_y, t)\), 
and fit the time dependence of the result using Eq.~\eqref{ft}. 
Comparing with the initial value of the imbalance, we find an estimate for the fit parameter $f_1$,
\begin{align}
   f_1 \sim \frac{\langle \tilde{n}(q_x, q_y, 0)\rangle}{\langle \tilde{n}(q=0) \rangle \nu}.
\end{align}

\bibliography{bibliography}

\end{document}